\documentclass{IEEEtran}
 \pdfoutput=1
\usepackage[T1]{fontenc}
\usepackage[utf8]{inputenc}
\usepackage{array}
\usepackage{float}
\usepackage{units}
\usepackage{amsmath}
\usepackage{amsthm}
\usepackage{amssymb}
\usepackage{stackrel}
\usepackage{graphicx}
\usepackage[unicode=true,
 bookmarks=true,bookmarksnumbered=true,bookmarksopen=true,bookmarksopenlevel=1,
 breaklinks=false,pdfborder={0 0 0},pdfborderstyle={},backref=false,colorlinks=false]
 {hyperref}
\hypersetup{pdftitle={Your Title},
 pdfauthor={Your Name},
 pdfpagelayout=OneColumn, pdfnewwindow=true, pdfstartview=XYZ, plainpages=false}

\makeatletter

\providecommand{\tabularnewline}{\\}
\floatstyle{ruled}
\newfloat{algorithm}{tbp}{loa}
\providecommand{\algorithmname}{Algorithm}
\floatname{algorithm}{\protect\algorithmname}

\theoremstyle{plain}
\newtheorem{thm}{\protect\theoremname}
\theoremstyle{remark}
\newtheorem{rem}[thm]{\protect\remarkname}
\theoremstyle{plain}
\newtheorem{lem}[thm]{\protect\lemmaname}

\usepackage[caption=false,font=normalsize,labelfont=sf,textfont=sf]{subfig}
\newtheorem{assumption}{Assumption}

\makeatother

\providecommand{\lemmaname}{Lemma}
\providecommand{\remarkname}{Remark}
\providecommand{\theoremname}{Theorem}

\begin{document}
\title{A Two-stage Multiband WiFi Sensing Scheme via Stochastic Particle-Based
Variational Bayesian Inference}
\author{Zhixiang~Hu,~An~Liu,~\IEEEmembership{Senior Member,~IEEE,} Yubo
Wan, \textit{Graduate Student Member, IEEE}, Tony Xiao Han and Minjian
Zhao,~\IEEEmembership{Member,~IEEE}\thanks{Zhixiang Hu, An Liu, Yubo Wan and Minjian Zhao are with the College
of Information Science and Electronic Engineering, Zhejiang University,
Hangzhou 310027, China, e-mail: \protect\href{http://anliu@zju.edu.cn}{anliu@zju.edu.cn}.}\thanks{Tony Xiao Han is with Huawei Technologies Co., Ltd, e-mail: \protect\href{http://tony.hanxiao@huawei.com}{tony.hanxiao@huawei.com}.}}
\maketitle
\begin{abstract}
Multiband fusion enhances WiFi sensing by jointly utilizing signals
from multiple non-contiguous frequency bands. However, in the multi-band
WiFi sensing signal model, there are many local optimums in the associated
likelihood function due to the existence of high frequency component
and phase distortion factors, posing challenges for high-accuracy
parameter estimation. To address this, we propose a two-stage scheme
equipped with different signal models derived from the original model,
where the first-stage coarse estimation is performed using a weighted
root MUSIC algorithm to narrow down the search range for the subsequent
stage, and the second-stage refined estimation utilizes a Bayesian
approach to avoid convergence to bad suboptimal solutions. Specifically,
we apply the block stochastic successive convex approximation (SSCA)
approach to derive a novel stochastic particle-based variational Bayesian
inference (SPVBI) algorithm in the refined stage. Unlike conventional
particle-based VBI (PVBI) that optimizes only particle probability
and incurs exponential per-iteration complexity with particle count,
our more flexible SPVBI algorithm optimizes both the position and
probability of each particle. Additionally, it utilizes block SSCA
to significantly improve sampling efficiency by averaging over iterations,
making it suitable for high-dimensional problems. Extensive simulations
demonstrate the superiority of our proposed algorithm over various
baseline methods.
\end{abstract}

\begin{IEEEkeywords}
WiFi sensing, multi-band fusion, variational Bayesian inference, stochastic
successive convex approximation.
\end{IEEEkeywords}

\section{Introduction}

\IEEEPARstart{F}{uture} Wi-Fi networks (e.g., Wi-Fi 7) are expected
to achieve integrated sensing and communication (ISAC) functionality
\cite{ISAC}, where Wi-Fi signals will be exploited to also provide
high-accuracy sensing services, such as indoor localization and tracking,
activity recognition, in-home digital health, etc \cite{Wifi7}. The
sensing performance generally depends on the bandwidth of the Wi-Fi
signal. However, increasing the bandwidth is not a scalable solution
due to limited spectrum resources. Besides, large bandwidth will bring
great pressure to signal acquisition, data transmission and processing,
which leads to a high hardware costs. Although there have been many
improvements in the super-resolution method of single-band signal
processing, it is still limited by the fixed signal bandwidth. Such
a situation has prompted the recent research interest in multiband
WiFi sensing, a technology that provides the potential to improves
the resolution and range accuracy by making use of the channel state
information (CSI) measurements across multiple non-contiguous frequency
bands.

Compared to single-band WiFi sensing, multiband sensing can reap extra
multiband gains, which consist of two parts: (i) Bandwidth-related
apertures gains resulted from multiband observations; (ii) Band gap
apertures gains brought by the difference of carrier frequency between
subbands \cite{ESPRIT2}. The different apertures and spectrum distribution
are shown in Fig. \ref{freq_band}, and details will be explained
in the following sections.

Despite the existence of the multiband gains, the multiband WiFi sensing
technology comes with unique challenges. One challenge is that the
high frequency carrier leads to a violent oscillation phenomenon of
the likelihood function, resulting in numerous bad local optimums
around the truth value of the delay. As a result, finding the global
optimum of the target parameter becomes intractable. The other challenge
is the phase misalignment of the received multiband signals caused
by hardware imperfections \cite{nonidealfoctor3,Decimeter-Level}.
Per-band channel impulse response (CIR) measurements are superimposed
with a random initial phase and a time synchronization error, which
are independent from one band to the other. If the phase calibration
process is not executed or the calibration precision is insufficient,
the sensing performance may be poor. Therefore, it is extremely difficult
to fully exploit this apertures gains especially in the presence of
phase distortion factors.

The existing multi-band sensing methods can be divided into three
categories: subspace based method, compressed sensing based method
and probabilistic inference based method. These related works are
summarized in Table. \ref{survey}. Unfortunately, these methods have
at least one of the limitations as listed below: (i) Underlying band
gap apertures gain from multiband is not fully exploited; (ii) Failure
to achieve low cost calibration of phase distortion factors. Consequently,
we propose a two-stage multiband Wi-Fi sensing scheme via stochastic
particle-based variational Bayesian inference (SPVBI) to overcome
these limitations. The main contributions are summarized as follows.
\begin{table*}[tbh]
\caption{\label{survey}A BRIEF SUMMARY OF THE EXISTING MULTIBAND SENSING METHODS.}

\centering{}%
\begin{tabular}{|>{\centering}m{1.8cm}|>{\centering}m{4.3cm}|>{\centering}m{1.1cm}|>{\centering}m{3.5cm}|>{\centering}m{3cm}|}
\hline
\multicolumn{1}{|>{\centering}p{1.8cm}|}{Category} & \multicolumn{1}{c|}{One sentence summary} & Publication & Pros & Cons\tabularnewline
\hline
\hline
\multicolumn{1}{|c|}{Subspace based} & Leverage the special structure in multiband signal subspace to achieve
estimation & \cite{ESPRIT2,ESPRIT1,WR-MUSIC,MUSIC_1,MUSICbase} & Low computational complexity & Sensitive to model error and noise interference\tabularnewline
\hline
Compressed sensing based & Exploit sparsity of multipath channel in delay domain to achieve estimation & \cite{Decimeter-Level,CS1,CS2,CS3} & No need to determine the number of scattering paths & (i) Require dense grids; (ii) High computational complexity\tabularnewline
\hline
Probabilistic Inference based & (i) Maximum likelihood estimation; (ii) Bayesian inference & \cite{SAGE,ML} & (i) Exploit prior information; (ii) High estimation accuracy & (i) Easy to fall into local optimum; (ii)Complicated algorithm design\tabularnewline
\hline
\end{tabular}
\end{table*}

\begin{itemize}
\item \textbf{Two-stage parameter estimation framework: }The conventional
two-stage framework adopted the same signal model but different estimation
algorithms. However, in our proposed two-stage estimation here, different
signal models derived from the original model are adopted, and associated
two-stage estimation algorithms are designed according to the statistical
characteristics of different models. Specifically, in the coarse estimation
stage, a simple but stable MUSIC-based coarse estimation is used to
narrow down the search range, so that numerous 'bad' local optimums
can be excluded from the global search space in the refined estimation
stage. In the refined estimation stage, we adopt a modified particle-based
VBI method, which can make full use of the intrinsic band gap aperture
gain in the multiband refined signal model to improve the estimation
accuracy.
\item \textbf{Stochastic particle-based variational Bayesian inference:}
The proposed SPVBI algorithm uses three innovative ideas to achieve
accurate posteriori estimation of target parameters with reduced complexity
and fast convergence speed. First, we adopt the particle-based approximation
to transform the multiple integral operation in VBI into multiple
weighted summation. Second, particle positions are updated in each
iteration to minimize the VBI objective function. Such improved degree
of freedom can further enhance the performance and accelerate the
convergence speed. Third, to avoid the exponential complexity with
the number of particles, we extend the SSCA approach in \cite{SSCA}
to block SSCA and apply it to improve the sampling efficiency of the
expectation operator in VBI iteration using the average-over-iteration
technique.
\item \textbf{Rigorous convergence analysis of SPVBI:} As an extension of
existing SSCA, SPVBI allows block-wise update by constructing a series
of parallel sub-surrogate functions, which poses new challenges for
theoretical analysis due to asynchronous update. Furthermore, with
the update of particles in each iteration, the distribution of random
states is no longer constant, but changes dynamically, which is challenging
for the theoretical proof. Despite these challenges, we prove that
SPVBI is guaranteed to converge to a stationary point of the VBI problem,
even though the number of samples used to calculate the expectation
in each iteration is fixed as a constant that does not increase with
the number of adopted particles.
\end{itemize}
The rest of the paper is organized as follows. In Section $\text{\mbox{II}}$,
we introduce the system model and the original estimation problem
in the multi-band WiFi sensing scenario, and propose a novel two-stage
signal model. In Section $\text{\mbox{III}}$, we present a two-stage
estimation framework. In Section $\text{\mbox{IV}}$, we introduce
the SPVBI algorithm for the second stage, along with an analysis of
its convergence. In Section $\text{\mbox{V}}$, we present numerical
simulations and a performance analysis. Finally, conclusions are presented
in Section $\text{\mbox{VI}}$.

Notations: $\delta(\cdot)$ denotes the Dirac’s delta function, $\mathop{{\rm var}}(\cdot)$
denotes the variance operator, and $\left\Vert \cdot\right\Vert $
denotes the Euclidean norm. For a matrix $\mathbf{A}$, $\mathbf{A}^{T}$
, $\mathbf{A}^{H}$, $\mathbf{A}^{-1}$, represent a transpose, complex
conjugate transpose and inverse, respectively. $\mathbb{E}_{z}[\cdot]$
denotes the expectation operator with respect to the random vector
$z$. $\boldsymbol{D}_{KL}\left[q\left\Vert p\right.\right]$ denotes
the Kullback-Leibler (KL) divergence of the probability distributions
$q$ and $p$. $\mathcal{N}(\mu,\Sigma)$ and $\mathcal{CN}(\mu,\Sigma)$
denotes Gaussian and complex Gaussian distribution with mean $\mu$
and covariance matrix $\Sigma$.

\section{System Model and Problem Formulation}

\subsection{System Model}

In the scene of multi-band WiFi sensing, we consider a single-input
single-output (SISO) system that uses OFDM training signals over $M$
frequency subbands to estimate range between the mobile node and Wi-Fi
device. The indoor radio propagation channel $h\left(t\right)$ between
transceivers is modeled as the sum of $K$ multipath components given
by
\begin{equation}
h_{m}\left(t\right)=\sum\limits _{k=1}^{K}\alpha_{k}e^{j\beta_{k}}\delta\left(t-\tau_{k}\right),\label{eq:CIR}
\end{equation}
where $m=1,2,\ldots M$ is the frequency band index, $\alpha_{k}e^{j\beta_{k}}$
is the complex gain carrying the amplitude and phase information of
a scattering path, and $\tau_{k}$ is the time delay of the $k$-th
path. Without loss of generality, we assume $\tau_{1}<\tau_{2}<\ldots<\tau_{K}$.
Therefore, $\tau_{1}$ represents the delay of line-of-sight (LoS)
path which is considered to be estimated for ranging.

Taking the Fourier transform of \eqref{eq:CIR}, the discrete frequency-domain
channel response can be expressed as
\begin{equation}
h_{m}\left(n\right)=\sum\limits _{k=1}^{K}\alpha_{k}e^{j\beta_{k}}e^{-j2\pi\left(f_{c,m}+nf_{s,m}\right)\tau_{k}},
\end{equation}
where $n=0,1,\ldots N_{m}-1$ denotes subcarrier index, $N_{m}$ denotes
the number of subcarriers in each band. $f_{c,m}$ and $f_{s,m}$
are the initial frequency and subcarrier spacing of $m$-th frequency
band, respectively.

Unfortunately, The pilot signals in a WiFi device are subject to the
phase distortion due to hardware imperfections as well as additive
noise. Specifically, due to the packet detection delay (PDD) and receiver
sampling frequency offset (SFO) \cite{nonidealfoctor3,hardware_imperfection2,nonidealfoctor1},
there is a timing synchronization error $\delta_{m}$ in the CSI measurements
of $m$-th frequency band. In addition, due to the hardware difference,
the CSI measurements are superimposed with a random initial phase
$\phi_{m}$ \cite{XuHuilin2}. These two imperfect factors are obstacles
for multi-band signal fusion and need to be calibrated.

After removing the known training signals, the discrete frequency-domain
received signal model during the period of a single OFDM symbol can
be formulated as \cite{ESPRIT2,CS1}
\begin{equation}
r_{m}^{\left(n\right)}=\sum\limits _{k=1}^{K}\alpha_{k}e^{j\beta_{k}}e^{-j2\pi\left(f_{c,m}+nf_{s,m}\right)\left(\tau_{k}+\delta_{m}\right)}e^{j\phi_{m}}+w_{m}^{\left(n\right)},\label{eq:original_model}
\end{equation}
$w_{m}^{\left(n\right)}$ denotes an additive white Gaussian noise
(AWGN) following the distribution $\mathcal{CN}\left(0,\eta_{w}^{2}\right)$.

\begin{figure}[htbp]
\begin{centering}
\textsf{\includegraphics[clip,scale=0.31]{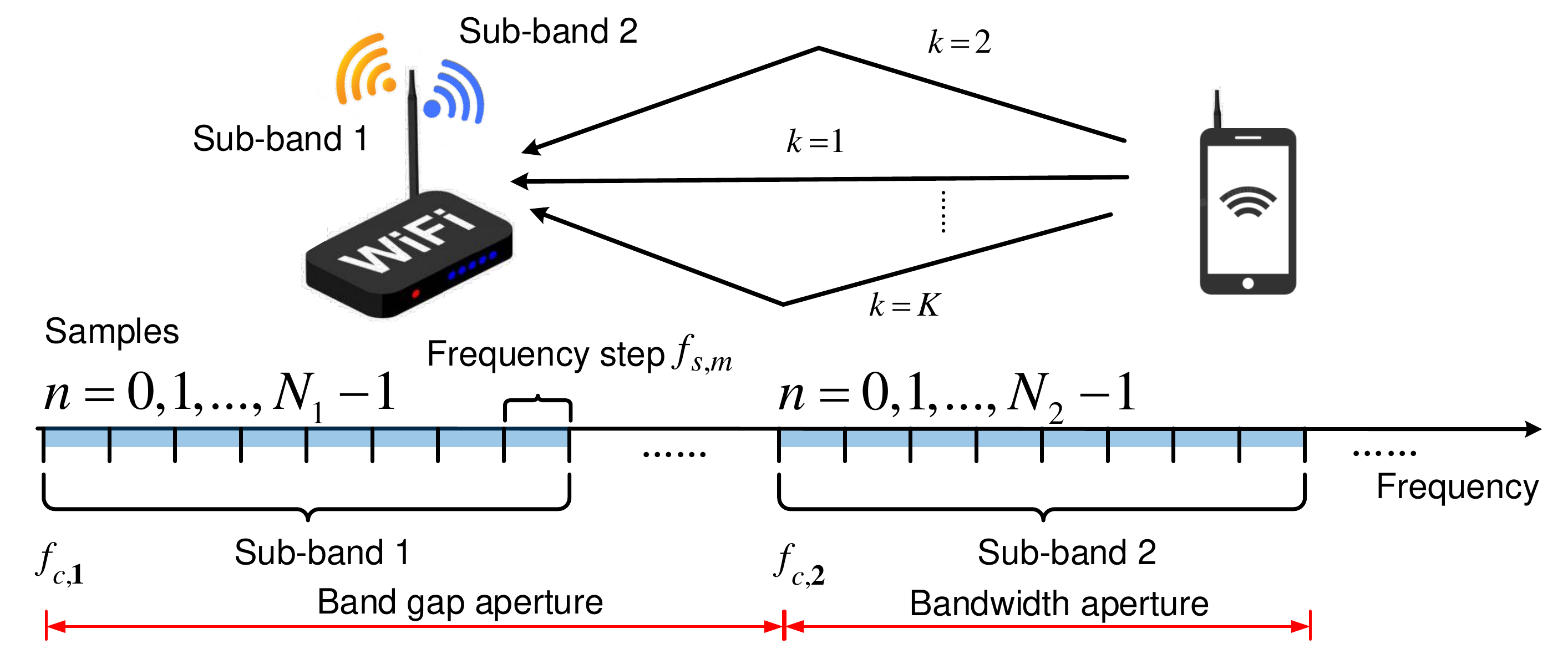}}
\par\end{centering}
\caption{\label{freq_band}An illustration of multi-band WiFi system and spectrum
distribution.}
\end{figure}

\subsection{Problem Formulation}

In multiband sensing problem, the variables to be estimated are denoted
as $\mathbf{\boldsymbol{\Lambda}_{ori}}{\rm =}\left[\alpha_{1},\ldots,\alpha_{K},\tau_{1},\ldots,\tau_{K},\beta_{1},\ldots,\beta_{K},\right.$
$\left.\phi_{1},\ldots,\phi_{M},\delta_{1},\ldots,\delta_{M}\right]^{T}$,
and frequency domain measurements of the received signal is vectorized
as $\boldsymbol{r}=\left[r_{1}^{\left(0\right)},r_{1}^{\left(1\right)},\ldots,r_{1}^{\left(N_{1}-1\right)},\right.$
$\left.\ldots,r_{M}^{\left(0\right)},r_{M}^{\left(1\right)},\ldots,r_{M}^{\left(N_{M}-1\right)}\right]^{T}$.

In the case of AWGN, the logarithmic likelihood function of the original
signal model \eqref{eq:original_model} can be written as follow
\begin{equation}
\begin{split} & \ln p\left(\boldsymbol{r}|\mathbf{\boldsymbol{\Lambda}_{ori}}\right)=\ln\prod\limits _{m=1}^{M}\prod\limits _{n=0}^{N_{m}-1}p(r_{m}^{\left(n\right)}|\mathbf{\boldsymbol{\Lambda}_{ori}})\\
 & =MN_{m}\ln\frac{1}{\sqrt{2\pi}\eta_{w}}-\sum_{m=1}^{M}\sum_{n=0}^{N_{m}-1}\frac{1}{2\eta_{w}^{2}}\left|r_{m}^{\left(n\right)}-s_{m}^{\left(n\right)}\left(\mathbf{\boldsymbol{\Lambda}_{ori}}\right)\right|^{2},
\end{split}
\label{eq:log_likelihood_orig}
\end{equation}
\begin{equation}
\begin{split}s_{m}^{\left(n\right)}\left(\mathbf{\boldsymbol{\Lambda}_{ori}}\right) & =\sum\limits _{k=1}^{K}\alpha_{k}e^{j\beta_{k}}e^{-j2\pi(f_{c,m}+nf_{s,m})\tau_{k}}\\
 & e^{j\phi_{m}}e^{-j2\pi\left(f_{c,m}+nf_{s,m}\right)\delta_{m}},
\end{split}
\label{eq:reconstruc_sig}
\end{equation}
where $s_{m}^{\left(n\right)}\left(\mathbf{\boldsymbol{\Lambda}_{ori}}\right)$
is the received signal reconstructed from the parameter $\mathbf{\boldsymbol{\Lambda}_{ori}}$.

Based on the above likelihood function and some given prior, we wish
to obtain the maximum a posteriori (MAP) estimate of the target parameters
by solving such a maximization problem:
\begin{equation}
\begin{split}\mathcal{P}_{\mathbf{\boldsymbol{\mathrm{ori}}}}:\mathbf{\left(\boldsymbol{\Lambda}_{ori}^{*}\right)_{MAP}=}\mathrm{arg}\mathop{\mathrm{max}}\limits _{\mathbf{\mathbf{\boldsymbol{\Lambda}_{ori}}}}\quad & \ln\left[p\left(\boldsymbol{r}|\mathbf{\boldsymbol{\Lambda}_{ori}}\right)p\left(\mathbf{\boldsymbol{\Lambda}_{ori}}\right)\right]\end{split}
,\label{eq:ori_MAP_problem}
\end{equation}
where the joint posteriori probability can be easily deduced from
the Bayes equation $p\left(\boldsymbol{\Lambda}|\boldsymbol{r}\right)\propto p\left(\boldsymbol{r}|\boldsymbol{\Lambda}\right)p\left(\boldsymbol{\Lambda}\right)$.

However, the global parameter estimation problem will be intractable
if we directly use the signal model \eqref{eq:original_model}. On
the one hand, the multi-dimensional parameters make the search space
of the estimation problem very huge. On the other hand, there are
many local optimums in the likelihood function \eqref{eq:log_likelihood_orig}.
Motivated by the these challenges, we next set up the novel two-stage
signal model and customize the associated two-stage global estimation
scheme.

\subsection{Two-stage Signal Model}

Prior to presenting the details of two-stage signal model, it is important
to note some features about the likelihood function in this problem.

\begin{figure}[htbp]
\begin{centering}
\textsf{\includegraphics[clip,scale=0.38]{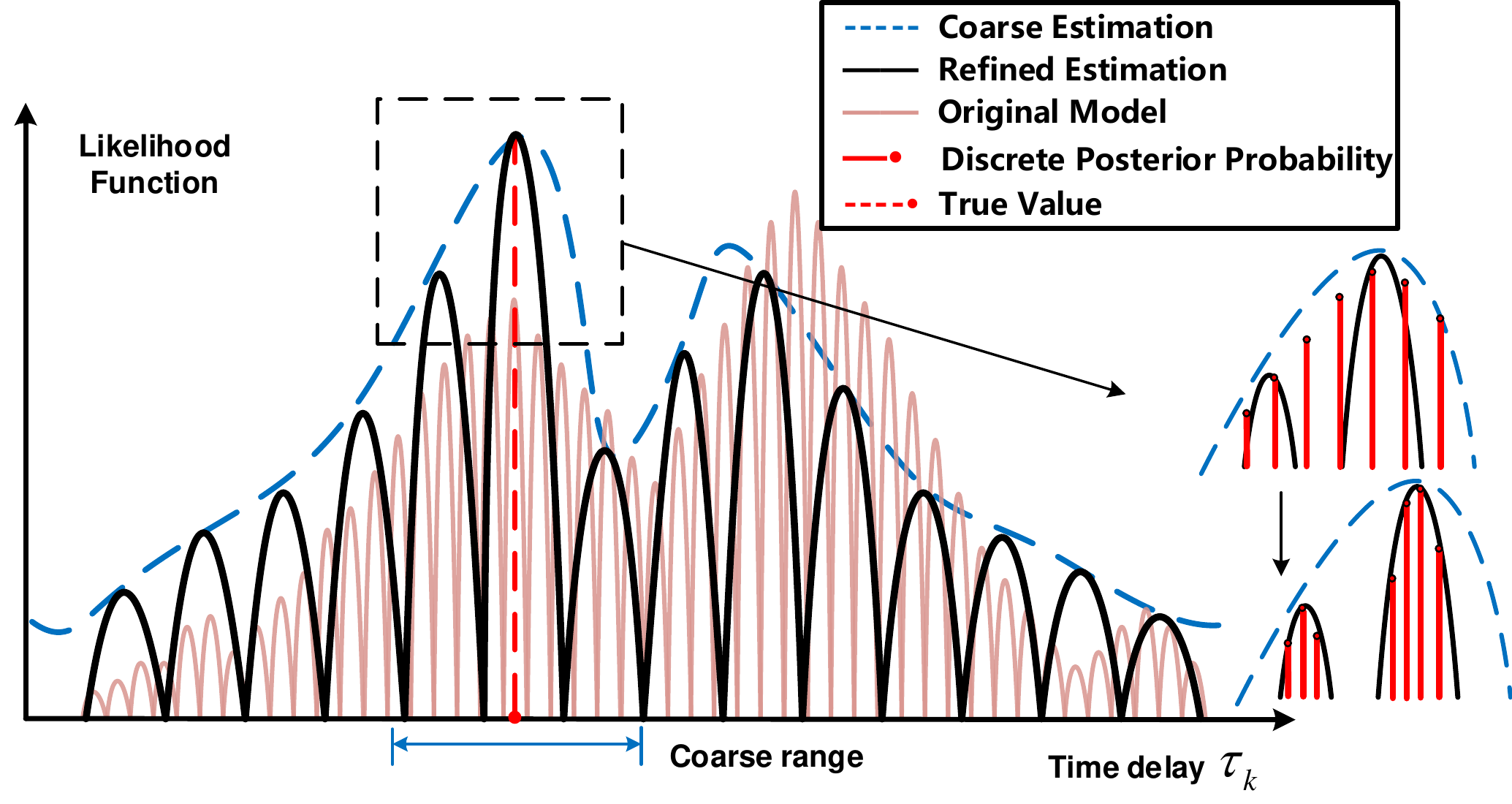}}
\par\end{centering}
\caption{\textsf{\label{oscillation_v5}}An illustration of oscillation phenomena
and particle approximation.}
\end{figure}

According to the above expression of the likelihood function \eqref{eq:log_likelihood_orig},
we can plot the spectrum of the likelihood function with respect to
time delay by fixing other variables. As shown by the solid pink line
in Fig. \ref{oscillation_v5}, the likelihood function fluctuates
periodically, with a sharp main lobe appearing at the location of
the true value point, accompanied by many oscillating side lobes.

As can be seen in \eqref{eq:log_likelihood_orig}, the period of this
oscillation phenomenon is affected by the frequency term $(f_{c,m}+nf_{s,m})$,
which is multiplied by the delay term $\tau_{k}$. Once the WiFi carrier
frequency reaches several to tens of GHz, this will inevitably lead
to violent oscillations of the likelihood function with a period of
nanoseconds. As a result, finding the global optimum becomes intractable.
Moreover, when the signal-to-noise ratio (SNR) is not high and there
are imperfect factors, the estimate may deviate to the locally optimal
side-lobe, resulting in a degradation of the estimation performance.

Therefore, we divide the target parameter estimation into two stages
equipped with different signal models derived from the original model
in \eqref{eq:original_model}. These two signal models have the structures
of bandwidth aperture and band gap aperture \cite{aperture} respectively,
which are presented in Fig. \ref{freq_band} and will be explained
below.

\textbf{(1) Coarse signal model with the bandwidth aperture $(nf_{s,m})$
structure:}
\begin{equation}
\begin{split}r_{m}^{\left(n\right)} & =\sum\limits _{k=1}^{K}\alpha_{k,m}^{'}e^{-j2\pi(nf_{s,m})\left(\tau_{k}+\delta_{m}\right)}+w_{m}^{\left(n\right)},\end{split}
\label{eq:coarse_model}
\end{equation}
where $\alpha_{k,m}^{'}=\alpha_{k}e^{j\beta_{k}}\cdot e^{j\phi_{m}}e^{-j2\pi f_{c,m}\left(\tau_{k}+\delta_{m}\right)}$.

In the coarse estimation stage, the carrier phase $e^{-j2\pi f_{c,m}\left(\tau_{k}+\delta_{m}\right)}$
and the random phase $e^{j\phi_{m}}$ of each band are absorbed into
the complex scalar $\alpha_{k}e^{j\beta_{k}}$, while the bandwidth
term $nf_{s,m}$ is retained, so that all sub-bands share a bandwidth-dependent
delay domain, which we call the bandwidth aperture.

As shown by the dotted blue line in Fig. \ref{oscillation_v5}, the
bandwidth aperture smoothes the likelihood function, so that the true
value can most likely be found in the peak region of the main lobe,
thus obtaining a relatively rough but stable estimate.

\textbf{(2) Refined signal model with the band gap aperture $(f_{c,m}^{'}+nf_{s,m})$
structure:}
\begin{equation}
\begin{split}r_{m}^{\left(n\right)} & =\sum\limits _{k=1}^{K}\alpha_{k}^{'}e^{j\beta_{k}^{'}}e^{-j2\pi(f_{c,m}^{'}+nf_{s,m})\tau_{k}}\\
 & e^{j\phi_{m}^{'}}e^{-j2\pi nf_{s,m}\delta_{m}}+w_{m}^{\left(n\right)},
\end{split}
\label{eq:refined_model}
\end{equation}
where, $\alpha_{k}^{'}e^{j\beta_{k}^{'}}=\alpha_{k}e^{j\beta_{k}}\cdot e^{j\tilde{\phi}_{1}}e^{-j2\pi f_{c,1}\tau_{k}}$,
$\phi_{m}^{'}=\tilde{\phi}_{m}-\tilde{\phi}_{1}$, $f_{c,m}^{'}=f_{c,m}-f_{c,1}$,
$\phi_{1}^{'}=0$, $f_{c,1}^{'}=0$.

In the stage of refined estimation, we first absorb the term $e^{-j2\pi f_{c,m}\delta_{m}}$
into the initial phase $e^{j\phi_{m}}$, namely $e^{j\tilde{\phi}_{m}}{\rm =}e^{j\left(\phi_{m}-2\pi f_{c,m}\delta_{m}\right)}$,
which also aims to reduce the oscillation degree of the likelihood
function in the estimation of $\delta_{m}$. Then, if we take the
first frequency band as a reference, the rewritten initial phase $e^{j\tilde{\phi}_{1}}$
and the carrier phase $e^{-j2\pi f_{c,1}\tau_{k}}$ can be absorbed
into the complex scalar $\alpha_{k}e^{j\beta_{k}}$ of each path,
and the residual band gap term $e^{-j2\pi(f_{c,m}-f_{c,1}+nf_{s,m})\tau_{k}}$
is retained in each sub-band signal, which is therefore called the
band gap aperture structure.

Fig. \ref{oscillation_v5} illustrates that the oscillation of likelihood
function associated with the refined signal model will not be too
violent, at the same time, the main lobe becomes sharper than that
of the coarse estimation model. In this case, we can exploit the multi-band
gain (i.e, the more sensitive phase rotation caused by the gap between
the carrier frequencies) to improve the performance.

\section{Two-stage Estimation Framework}

Based on the two-stage signal model, the two-stage estimation framework
is depicted as follow.

\subsection{Coarse Estimation\label{subsec:Coarse-Estimation}}

In the coarse estimation stage, we first need to determine the number
of scattering paths $K$. To the best of our knowledge, Akaike Information
Criterion (AIC) \cite{AIC} or the Minimum Description Length (MDL)
\cite{MDLprinciple} are both efficient methods that generally work
well, which will not be described here for conciseness. Then, we use
the weighted root-MUSIC (WR-MUSIC) \cite{WR-MUSIC,rao1989root-MUSIC}
algorithm to roughly estimate the delay $\tau_{k}$. The coarse estimate
signal model \eqref{eq:coarse_model} can be written in the following
form:
\begin{equation}
\boldsymbol{\mathbf{Y}}_{m}=\boldsymbol{\mathbf{X}}_{m}\boldsymbol{\mathbf{A}}_{m}+\boldsymbol{w}_{m},
\end{equation}
where
\begin{equation}
\begin{cases}
\boldsymbol{\mathbf{Y}}_{m}{\rm =}\left[r_{m}^{\left(0\right)},r_{m}^{\left(1\right)},\ldots,r_{m}^{\left(N_{m}-1\right)}\right]^{T},\\
\boldsymbol{\mathbf{X}}_{m}(\boldsymbol{\mathbf{\tau}}){\rm =}\left[\boldsymbol{\mathbf{x}}(\tau_{1}+\delta_{m}),\boldsymbol{\mathbf{x}}(\tau_{2}+\delta_{m}),\cdots,\mathbf{\boldsymbol{x}}(\tau_{K}+\delta_{m})\right],\\
\boldsymbol{\mathbf{x}}(\mathring{\tau}){\rm =}[1,e^{-j2\pi f_{s,m}\mathring{\tau}},\cdots,e^{-j2\pi(N_{m}-1)f_{s,m}\mathring{\tau}}]^{T},\\
\mathbf{\boldsymbol{A}}_{m}{\rm =}[\alpha_{1,m}^{'},\alpha_{2,m}^{'},\cdots,\alpha_{K,m}^{'}]^{T},\\
\boldsymbol{w}_{m}=[w_{m}^{\left(0\right)},w_{m}^{\left(1\right)},\cdots,w_{m}^{\left(N_{m}-1\right)}]^{T},
\end{cases}
\end{equation}
and $\boldsymbol{\mathbf{\tau}}=\left[\tau_{1},...,\tau_{K}\right]^{T}$.
Since the coarse estimation aims to simply reduce the search range
with low complexity and doesn't require high accuracy, it's reasonable
to use such a simplified form even if it may introduce some model
distortion.

Next according to the WR-MUSIC algorithm, we can construct a Hankel
matrix for subspace decomposition of the received signal \cite{subspace},
in the following form:
\begin{equation}
\boldsymbol{H}_{m}=\left[\begin{array}{cccc}
r_{m}^{\left(0\right)} & r_{m}^{\left(1\right)} & \cdots & r_{m}^{\left(L-1\right)}\\
r_{m}^{\left(1\right)} & r_{m}^{\left(2\right)} & \cdots & r_{m}^{\left(L\right)}\\
\vdots & \vdots & \ddots & \vdots\\
r_{m}^{\left(N_{m}-L\right)} & r_{m}^{\left(N_{m}-L+1\right)} & \cdots & r_{m}^{\left(N_{m}-1\right)}
\end{array}\right],
\end{equation}
where $L$ is the length of correlation window, empirically taken
as $\nicefrac{N_{m}}{3}$ \cite{UWB}.

After that, we apply the eigenvalue decomposition to the Hankel matrix
of the signal, $\boldsymbol{H}_{m}=\boldsymbol{U}_{m}\boldsymbol{D}_{m}\boldsymbol{V}_{m}^{H}$.
The eigenspace composed of the eigenvectors corresponding to the largest
$K$ (i.e. the number of scattering paths) eigenvalues is called the
signal subspace and is denoted as $\ensuremath{\boldsymbol{\mathbf{S}}_{signal}}$.
The eigenspace composed of the eigenvectors corresponding to the remaining
$(N_{m}-K)$ eigenvalues is called the noise subspace and denoted
as $\ensuremath{\boldsymbol{\mathbf{S}}_{noise}}$.

Essentially, compared with the conventional MUSIC algorithm, the root
MUSIC algorithm \cite{rao1989root-MUSIC} is to apply the polynomial
root-finding method to replace the spectral search of zeros. Define
the polynomial: $f_{l}\left(z\right)=\mathbf{u}_{l}^{H}\mathbf{p}\left(z\right),l=K+1,\ldots N_{m}$,
where $\mathbf{u}_{l}^{H}$ is the $l$-th eigenvector of noise subspace
$\ensuremath{\boldsymbol{\mathbf{S}}_{noise}}$, $\mathbf{p}\left(z\right)=[1,z,\cdots,z^{N_{m}-1}]^{T}$,
and $z=e^{-j2\pi f_{s,m}\tau}$. In order to utilize all noise eigenvectors,
we wish to find the zeros of the following polynomials:
\begin{equation}
f\left(z\right)=\mathbf{p}^{H}\left(z\right)\mathbf{S}_{noise}\mathbf{S}_{noise}^{H}\mathbf{p}\left(z\right).\label{eq:7}
\end{equation}

We rewrite \eqref{eq:7} to get the polynomial in terms of $z$ as:
\begin{equation}
f\left(z\right)=z^{N_{m}-1}\mathbf{p}^{T}\left(z^{-1}\right)\mathbf{S}_{noise}\mathbf{S}_{noise}^{H}\mathbf{p}\left(z\right).
\end{equation}

Find the roots of the above polynomial, wherein the $K$ roots in
the unit circle whose moduli are closest to $1$ contain the information
about delay. Denote those roots as $p_{k,m},k=1,2\ldots K$, then
the coarse estimate of the delay can be obtained: $\hat{\tau}_{k,m}=\frac{arg\left(p_{k,m}\right)}{-2\pi f_{s,m}}$.
We can further improve the delay estimation accuracy by combining
the results of different frequency band according to the Cramér-Rao
bound (CRB) analysis \cite{WR-MUSIC}:
\begin{equation}
\hat{\tau}_{k}=\frac{\sum_{m=1}^{M}SNR_{m}\cdot B_{m}\cdot\left(f_{c,m}^{2}+B_{m}^{2}/12\right)\cdot\hat{\tau}_{k,m}}{\sum_{m=1}^{M}SNR_{m}\cdot B_{m}\cdot\left(f_{c,m}^{2}+B_{m}^{2}/12\right)}.
\end{equation}

Also, the estimate of $\alpha_{k,m}^{'}$ can be obtained by the least
square (LS) method:
\begin{equation}
\boldsymbol{\mathbf{\hat{A}}}_{m}=\left(\boldsymbol{\mathbf{X}}_{m}^{H}(\hat{\boldsymbol{\tau}})\boldsymbol{\mathbf{X}}_{m}(\hat{\boldsymbol{\tau}})\right)^{-1}\boldsymbol{\mathbf{X}}_{m}^{H}(\hat{\boldsymbol{\tau}})\boldsymbol{\mathbf{Y}}_{m}.
\end{equation}
where $\hat{\boldsymbol{\tau}}=\left[\hat{\tau}_{1},...,\hat{\tau}_{K}\right]$,
$\boldsymbol{\mathbf{\hat{A}}}_{m}=[\hat{\alpha}_{1,m}^{'},\hat{\alpha}_{2,m}^{'},\cdots,\hat{\alpha}_{K,m}^{'}]^{T}$.
Then, the signal can be written as an all-pole model \cite{UWB}:
\begin{equation}
r_{m}^{\left(n\right)}=\sum\limits _{k=1}^{K}\hat{\alpha}_{k,m}^{'}\left(e^{-j2\pi f_{s,m}\hat{\tau}_{k,m}}\right)^{n}=\sum\limits _{k=1}^{K}\hat{\alpha}_{k,m}^{'}p_{k,m}^{n}.\label{eq:model_allpoles}
\end{equation}

By comparing the terms in equations \eqref{eq:coarse_model} and \eqref{eq:model_allpoles},
the difference between the random initial phases of the two bands
can be estimated as follows:
\begin{equation}
\begin{split}\hat{\phi}_{m}^{'}= & \hat{\phi}_{m}-\hat{\phi}_{1}=\frac{1}{K}\sum\limits _{k=1}^{K}\left[arg\left(\hat{\alpha}_{k,m}^{'}\right)-arg\left(\hat{\alpha}_{1,m}^{'}\right)\right.\\
 & \left.-2\pi f_{c,1}\left(\hat{\tau}_{k,1}\right)+2\pi f_{c,m}\left(\hat{\tau}_{k\text{,m}}\right)\right].
\end{split}
\end{equation}

Then we can also get
\begin{equation}
\hat{\delta}_{m}=\frac{1}{K}\sum\limits _{k=1}^{K}\left[\frac{arg\left(p_{k,m}\right)}{-2\pi f_{s,m}}-\hat{\tau}_{k}\right].
\end{equation}

Therefore, in the coarse estimation stage, we can obtain coarse estimate
of delay $\hat{\tau}_{k}$, complex gain amplitude $\left\Vert \widehat{\alpha}_{k}\right\Vert $
(i.e. $\frac{\sum_{m=1}^{M}SNR_{m}\cdot B_{m}\cdot\left(f_{c,m}^{2}+B_{m}^{2}/12\right)\cdot\left\Vert \hat{\alpha}_{k,m}^{'}\right\Vert }{\sum_{m=1}^{M}SNR_{m}\cdot B_{m}\cdot\left(f_{c,m}^{2}+B_{m}^{2}/12\right)}$),
difference between the random initial phases $\hat{\phi}_{m}^{'}$
and timing synchronization error $\hat{\delta}_{m}$ to serve the
refined estimation.

In addition, coarse estimates and reasonable prior assumptions together
provide prior information for the subsequent stage. As mentioned earlier,
the truth value $\tau_{k}$ will most likely fall in the neighborhood
of the coarse estimate $[\hat{\tau}_{k}-\Delta\hat{\tau}_{k}/2,\hat{\tau}_{k}+\Delta\hat{\tau}_{k}/2]$.
The interval of refined estimation $\Delta\hat{\tau}_{k}$ can be
determined according to the empirical error of the first stage, or
several times of the square root of the CRB calculated based on the
coarse estimate. Although a relatively accurate estimate of $\left\Vert \widehat{\alpha}_{k}\right\Vert $
is obtained in the first stage, its phase is still unknown. It can
be assumed that the phase is uniformly distributed from $0$ to $2\pi$.
The prior probability distributions for $\phi_{m}^{'}$ and $\beta_{k}^{'}$
are similar. Since the timing synchronization error $\delta_{m}$
is often within a small interval, it can be assumed that it follows
a prior distribution $\mathcal{N}\left(0,\sigma^{2}\right)$ with
a relatively small variance $\sigma^{2}$ \cite{nonidealfoctor3}.

\subsection{Refined Estimation}

Correspondingly, variables to be estimated in the refined stage are
adjusted as $\mathbf{\boldsymbol{\Lambda}}{\rm =}\left[\alpha_{1}^{'},\ldots,\alpha_{K}^{'},\tau_{1},\ldots,\tau_{K},\right.$
$\left.\beta_{1}^{'},\ldots,\beta_{K}^{'},\phi_{2}^{'},\ldots,\phi_{M}^{'},\delta_{1},\ldots,\delta_{M}\right]^{T}$
and the total number of variables is denoted as $\left|\boldsymbol{\Lambda}\right|=J$.
In the logarithmic likelihood function \eqref{eq:log_likelihood_orig},
$s_{m}^{\left(n\right)}$ is also adjusted to
\begin{equation}
\begin{split}s_{m}^{\left(n\right)}\left(\boldsymbol{\Lambda}\right) & =\sum\limits _{k=1}^{K}\alpha_{k}^{'}e^{j\beta_{k}^{'}}e^{-j2\pi(f_{_{c,m}}^{'}+nf_{s,m})\tau_{k}}\\
 & e^{j\phi_{m}^{'}}e^{-j2\pi\left(nf_{s,m}\right)\delta_{m}}.
\end{split}
\end{equation}

In summary, we provide a two-stage estimation scheme as shown in Algorithm
1, which divides the estimation process into coarse estimation stage
and refined estimation stage.

\begin{algorithm}[tbh]
\caption{Two-stage Estimation Scheme}

\textbf{Input}: received signals $\boldsymbol{r}$, multi-band frequency
settings.

\textbf{Stage 1: Coarse Estimation}

$\quad\quad$Construct signal model with the bandwidth

$\quad\quad$aperture structure \eqref{eq:coarse_model};

$\quad\quad$Perform WR-MUSIC and LS method;

$\quad\quad$Provide coarse estimates and prior intervals for Stage
2.

\textbf{Stage 2: Refined Estimation}

$\quad\quad$Construct signal model with the band gap

$\quad\quad$aperture structure \eqref{eq:refined_model};

$\quad\quad$Perform SPVBI algorithm to obtain the approximate

$\quad\quad$marginal posteriori probability.

\textbf{Output}: Obtain the MAP/MMSE estimates of each variable based
on approximate marginal posteriori probability.
\end{algorithm}

In the coarse estimation stage, WR-MUSIC algorithm and LS method are
used to obtain the rough estimates, providing prior information for
subsequent estimation and narrowing the search interval. In the next
section, we shall propose a SPVBI algorithm to find the approximate
marginal posteriori of the target parameters $\boldsymbol{\Lambda}$
for the refined estimation stage.

\section{Stochastic Particle-Based Variational Bayesian Inference}

\subsection{Problem Formulation for SPVBI}

In the original MAP estimation problem \eqref{eq:ori_MAP_problem},
we need to derive marginal posteriori probability distribution $p\left(\mathbf{\boldsymbol{\Lambda}}_{j}|\boldsymbol{r}\right)$
to perform posteriori inference on the target parameters, where $\mathbf{\mathbf{\boldsymbol{\Lambda}}}_{j}$
represents the $j$-th variable to be estimated in $\mathbf{\boldsymbol{\Lambda}}$.
In general, it is intractable to get a closed-form solution for $p\left(\mathbf{\boldsymbol{\Lambda}}_{j}|\boldsymbol{r}\right)$,
since it is necessary to integrate all the other variables except
$\mathbf{\boldsymbol{\Lambda}}_{j}$ \cite{tutorial_VBI} (i.e., multiple
expectation operation). To deal with this, many existing works make
prior assumptions, such as assuming that the distribution of these
variables comes from certain special distribution families \cite{Beal2003}
or meets the conjugation condition. However, these assumptions are
often not accurate enough and can be subjective.

In \cite{VBI+IS}, the authors proposed a particle-based VBI (PVBI)
algorithm to approximate the expectation calculation using importance
sampling method \cite{particle-assisted}, which is a type of Monte
Carlo method. By iteratively updating the weight of particles, the
discrete distribution composed of particles $q\left(\mathbf{\mathbf{\boldsymbol{\Lambda}}}_{j}\right)$
can gradually approach the true posteriori probability distribution
$p\left(\mathbf{\boldsymbol{\Lambda}}_{j}|\boldsymbol{r}\right)$.

The discrete variational posteriori probability $q\left(\mathbf{\mathbf{\boldsymbol{\Lambda}}}_{j}\right)$
can be written as the weighted sum of $N_{p}$ discrete particles
\cite{VBI+IS}:
\begin{equation}
q\left(\mathbf{\mathbf{\boldsymbol{\Lambda}}}_{j};\mathbf{x}_{j},\mathbf{y}_{j}\right){\rm =}\sum\limits _{p=1}^{N_{p}}y_{j,p}\delta\left(\mathbf{\boldsymbol{\Lambda}}_{j}-x_{j,p}\right),\label{eq:weighted sum-1}
\end{equation}
where $\mathbf{x}_{j}=\left[x_{j,1},x_{j,2},...,x_{j,N_{p}}\right]^{T}$
and $\mathbf{y}_{j}=\left[y_{j,1},y_{j,2},...,y_{j,N_{p}}\right]^{T}$
are the positions and weights of the particles, respectively. Furthermore,
according to the mean field assumption \cite{parisi1988meanfield},
the approximate posteriori probability of each variable can be assumed
to be independent of each other, $q\left(\boldsymbol{\Lambda};\mathbf{x},\mathbf{y}\right)=\stackrel[j=1]{J}{\prod}q\left(\boldsymbol{\Lambda}_{j};\mathbf{x}_{j},\mathbf{y}_{j}\right)$,
where $\mathbf{x}=\left[\mathbf{x}_{1};\mathbf{x}_{2};...;\mathbf{x}_{J}\right]$
and $\mathbf{y}=\left[\mathbf{y}_{1};\mathbf{y}_{2};...;\mathbf{y}_{J}\right]$.
Next, the positions and weights $\mathbf{x},\mathbf{y}$ are chosen
to minimize the Kullback-Leibler (KL) divergence between the variational
probability distribution $q\left(\mathbf{\boldsymbol{\Lambda}};\mathbf{x},\mathbf{y}\right)$
and the real posteriori probability distribution $p\left(\mathbf{\boldsymbol{\Lambda}}|\boldsymbol{r}\right)$
\cite{tutorial_VBI}, which is defined as
\begin{equation}
\begin{split}\boldsymbol{D}_{KL}\left[q\left\Vert p\right.\right] & {\rm =}\int q\left(\mathbf{\boldsymbol{\Lambda}};\mathbf{x},\mathbf{y}\right)\ln\frac{q\left(\mathbf{\mathbf{\boldsymbol{\Lambda}}};\mathbf{x},\mathbf{y}\right)}{p\left(\mathbf{\boldsymbol{\Lambda}}|\boldsymbol{r}\right)}d\mathbf{\boldsymbol{\Lambda}}\\
 & {\rm =}\int q\left(\mathbf{\boldsymbol{\Lambda}};\mathbf{x},\mathbf{y}\right)\ln\frac{q\left(\mathbf{\mathbf{\boldsymbol{\Lambda}}};\mathbf{x},\mathbf{y}\right)p\left(\boldsymbol{r}\right)}{p\left(\boldsymbol{r}|\mathbf{\boldsymbol{\Lambda}}\right)p\left(\mathbf{\boldsymbol{\Lambda}}\right)}d\mathbf{\mathbf{\boldsymbol{\Lambda}}}.
\end{split}
\end{equation}
Considering that $p\left(\boldsymbol{r}\right)$ is a constant independent
of $q\left(\mathbf{\boldsymbol{\Lambda}};\mathbf{x},\mathbf{y}\right)$,
minimizing the KL divergence is equivalent to solving the following
optimization problem:
\[
\begin{split}\mathcal{P}:\mathop{\min}\limits _{\mathbf{x},\mathbf{y}}\quad & \boldsymbol{L}\left(\mathbf{x},\mathbf{y}\right)\triangleq\int q\left(\boldsymbol{\Lambda};\mathbf{x},\mathbf{y}\right)\ln\frac{q\left(\boldsymbol{\Lambda};\mathbf{x},\mathbf{y}\right)}{p\left(\boldsymbol{r}|\boldsymbol{\Lambda}\right)p\left(\boldsymbol{\Lambda}\right)}d\boldsymbol{\Lambda}\\
\text{s.t.}\quad & \sum\limits _{p=1}^{N_{p}}y_{j,p}=1,\quad\epsilon\leq y_{j,p}\leq1,\quad\forall j,p,\\
 & \hat{\boldsymbol{\Lambda}}_{j}-\Delta\hat{\boldsymbol{\Lambda}}_{j}/2\le x_{j,p}\le\hat{\boldsymbol{\Lambda}}_{j}+\Delta\hat{\boldsymbol{\Lambda}}_{j}/2,\quad\forall j,p,
\end{split}
\]
where $p=1,2,\ldots N_{p}$ and $j=1,2,\ldots J$ are the particle
index and variables index, respectively. $\boldsymbol{L}\left(\mathbf{x},\mathbf{y}\right)$
denotes the objective function and $\epsilon>0$ is a small number.
The normalized weight $y_{j,p}$ represents the probability that the
particle is located at position $x_{j,p}$. $\hat{\boldsymbol{\Lambda}}_{j}$
is the coarse estimation obtained in the first stage and $\Delta\hat{\boldsymbol{\Lambda}}_{j}$
is the range of the estimation error of coarse estimation. The truth
value is highly likely to be located in the prior interval $\left[\hat{\boldsymbol{\Lambda}}_{j}-\Delta\hat{\boldsymbol{\Lambda}}_{j}/2,\hat{\boldsymbol{\Lambda}}_{j}+\Delta\hat{\boldsymbol{\Lambda}}_{j}/2\right]$,
so the particle position is searched wherein to accelerate the convergence
rate. Note that it does not make sense to generate particles with
very small probabilities in approximate posteriori since these particles
contribute very little to the estimator. Therefore, we restrict the
probability of each particle $y_{j,p}$ to be larger than a small
number $\epsilon$.

The solution of problem $\mathcal{P}$ refers to the marginal posteriori
distribution of each variable, from which we can obtain the best posteriori
estimate of the target parameter. In the following sections, we will
discuss how to solve this problem.
\begin{rem}
In conventional PVBI, the position of particles are not updated. Besides,
a large number of particles are required to overcome the instability
caused by initial random sampling, and ensure that the estimation
locally converges to a 'good' stationary point. As such, the complexity
will rocket as the number of variables and particles increases. Motivated
by this observcation, we optimize the particle position as well and
further design an SPVBI algorithm to solve the new problem with much
lower per-iteration complexity than the conventional PVBI algorithm.
Adding the optimization of particle positions $\mathbf{x}$ can improve
the effectiveness of characterizing the target distribution by discrete
particles, and avoid the estimation result falling into the local
optimum due to poor initial sampling. Furthermore, updating particle
position can reduce the number of required particles, and thus effectively
reduce the computational overhead, which will be discussed in detail
in the following sections.
\end{rem}

\subsection{SPVBI Algorithm Design based on Block SSCA\label{subsec:SPVBI-Algorithm-Design}}

Although updating particle positions can speed up convergence, multiple
integration in the objective function $\boldsymbol{L}\left(\mathbf{x},\mathbf{y}\right)$
of $\mathcal{P}$ is still intractable. To solve this problem, we
propose a stochastic particle-based variational Bayesian inference
algorithm based on the block SSCA to find stationary points of $\mathcal{P}$
with lower computational complexity.

Specifically, we divided the optimization variables into $2J$ blocks
$\mathbf{x}_{1}$, $\mathbf{y}_{1}$, $\mathbf{x}_{2}$, $\mathbf{y}_{2}$,
... ,$\mathbf{x}_{J}$, $\mathbf{y}_{J}$. Starting from an initial
point $\mathbf{x}^{\left(0\right)},\mathbf{y}^{\left(0\right)}$,
the SPVBI algorithm alternatively optimizing each block until convergence.
Let $\mathbf{x}_{j}^{\left(t\right)},\mathbf{y}_{j}^{\left(t\right)}$
and $\mathbf{x}_{j}^{\left(t+1\right)},\mathbf{y}_{j}^{\left(t+1\right)}$
denote the blocks $\mathbf{x}_{j},\mathbf{y}_{j}$ before and after
the update in the $t$-th iteration, respectively. Then in the $t$-th
iteration, the $\left(2j-1\right)$-th block $\mathbf{x}_{j}$ is
updated by solving the following subproblem:
\[
\begin{split}\mathcal{P}_{x_{j}}:\mathop{\min}\limits _{\mathbf{x}_{j}}\quad & \boldsymbol{L}_{j}^{\left(t\right)}\left(\mathbf{x}_{j},\mathbf{y}_{j}^{\left(t\right)}\right)\\
\text{s.t.}\quad & \hat{\boldsymbol{\Lambda}}_{j}-\Delta\hat{\boldsymbol{\Lambda}}_{j}/2\le x_{j,p}\le\hat{\boldsymbol{\Lambda}}_{j}+\Delta\hat{\boldsymbol{\Lambda}}_{j}/2,\forall p
\end{split}
\]
where $\boldsymbol{L}_{j}^{\left(t\right)}\left(\mathbf{x}_{j},\mathbf{y}_{j}\right)=\boldsymbol{L}\left(\mathbf{x}^{\left(t,j\right)},\mathbf{y}^{\left(t,j\right)}\right)$
with $\mathbf{x}^{\left(t,j\right)}=\left[\mathbf{x}_{1}^{\left(t+1\right)}\right.$
$\left.;...;\mathbf{x}_{j-1}^{\left(t+1\right)};\mathbf{x}_{j};\mathbf{x}_{j+1}^{\left(t\right)};...;\mathbf{x}_{J}^{\left(t\right)}\right]$
and $\mathbf{y}^{\left(t,j\right)}=\left[\mathbf{y}_{1}^{\left(t+1\right)};...;\mathbf{y}_{j-1}^{\left(t+1\right)}\right.$
$\left.;\mathbf{y}_{j};\mathbf{y}_{j+1}^{\left(t\right)};...;\mathbf{y}_{J}^{\left(t\right)}\right]$.
In other words, $\boldsymbol{L}_{j}^{\left(t\right)}\left(\mathbf{x}_{j},\mathbf{y}_{j}\right)$
is the objective function $\boldsymbol{L}\left(\mathbf{x},\mathbf{y}\right)$
when fixing all other variables as the latest iterate and only treating
$\mathbf{x}_{j},\mathbf{y}_{j}$ as variables. It can be shown that
\begin{align}
 & \boldsymbol{L}_{j}^{\left(t\right)}\left(\mathbf{x}_{j},\mathbf{y}_{j}\right)=\sum\limits _{p=1}^{N_{p}}y_{j,p}\ln y_{j,p}-\sum\limits _{p=1}^{N_{p}}y_{j,p}\left[\ln p\left(x_{j,p}\right)+\right.\nonumber \\
 & \sum\limits _{p_{1}=1}^{N_{p}}\cdots\sum\limits _{p_{j-1}=1}^{N_{p}}\sum\limits _{p_{j+1}=1}^{N_{p}}\cdots\sum\limits _{p_{J}=1}^{N_{p}}\widetilde{y}_{\sim j,p_{\sim j}}\left.\ln p\left(\boldsymbol{r}\left|\widetilde{x}_{\sim j,p_{\sim j}},x_{j,p}\right.\right)\right],\label{eq:Lj-1}
\end{align}
where $\widetilde{y}_{\sim j,p_{\sim j}}=\underset{i\neq j}{\prod}y_{i,p_{i}}$
and $\widetilde{x}_{\sim j,p_{\sim j}}=\left\{ x_{i,p_{i}}\right\} _{i\neq j}$,
which involves a summation of $N_{p}^{J-1}$ terms. As such, the exponential
complexity of directly solving $\mathcal{P}_{x_{j}}$ is unacceptable.

To overcome this challenge, we reformulate $\mathcal{P}_{x_{j}}$
to a stochastic optimization problem as
\[
\begin{split}\mathcal{P}_{x_{j}}:\mathop{\min}\limits _{\mathbf{x}_{j}}\quad & \boldsymbol{L}_{j}^{\left(t\right)}\left(\mathbf{x}_{j},\mathbf{y}_{j}^{\left(t\right)}\right)\triangleq{\rm \mathbb{E}}_{\boldsymbol{\Lambda}_{\sim j}}\left[g_{j}^{\left(t\right)}\left(\mathbf{x}_{j},\mathbf{y}_{j}^{\left(t\right)};\boldsymbol{\Lambda}_{\sim j}\right)\right]\\
\text{s.t.}\quad & \hat{\boldsymbol{\Lambda}}_{j}-\Delta\hat{\boldsymbol{\Lambda}}_{j}/2\le x_{j,p}\le\hat{\boldsymbol{\Lambda}}_{j}+\Delta\hat{\boldsymbol{\Lambda}}_{j}/2,\forall p,
\end{split}
\]
where $\boldsymbol{\Lambda}_{\sim j}$ represents all the other variables
except $\boldsymbol{\Lambda}_{j}$ and ${\rm \mathbb{E}}_{\boldsymbol{\Lambda}_{\sim j}}\left[\cdot\right]$
represents the expectation operator over the probability distribution
of variable $\boldsymbol{\Lambda}_{\sim j}$, and
\begin{align}
 & g_{j}^{\left(t\right)}\left(\mathbf{x}_{j},\mathbf{y}_{j};\boldsymbol{\Lambda}_{\sim j}\right){\rm =}\sum\limits _{p=1}^{N_{p}}y_{j,p}\ln y_{j,p}-\sum\limits _{p=1}^{N_{p}}y_{j,p}\nonumber \\
 & \times\left[\ln p\left(x_{j,p}\right)+\ln p\left(\boldsymbol{r}\left|\boldsymbol{\Lambda}_{{\rm \sim}j},x_{j,p}\right.\right)\right].
\end{align}

\begin{figure}[htbp]
\begin{centering}
\textsf{\includegraphics[scale=0.6]{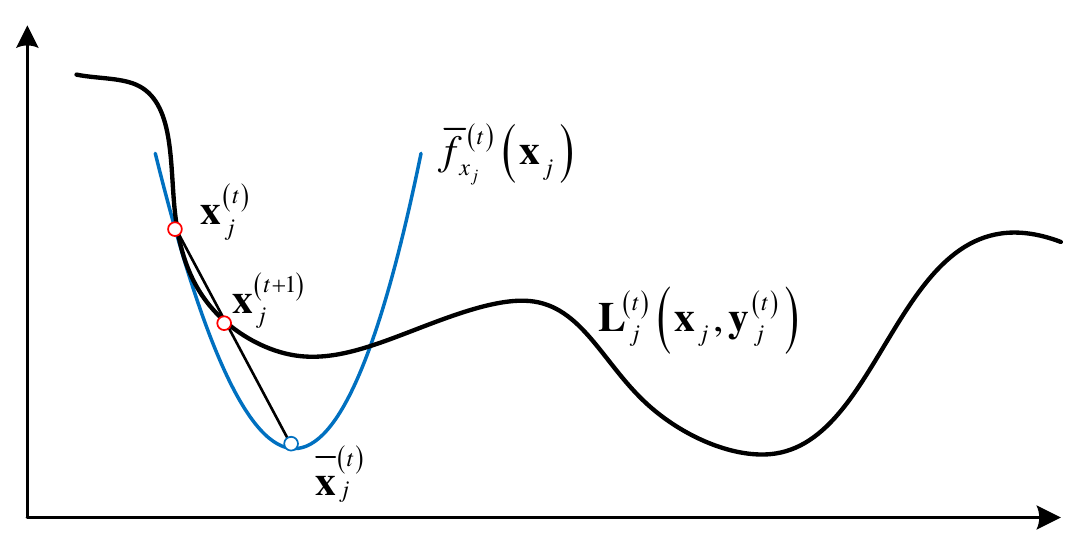}}
\par\end{centering}
\caption{\textsf{\label{SCA}}An illustration of stochastic successive convex
approximation.}
\end{figure}

Then following the idea of SSCA, we replace the objective function
$\boldsymbol{L}_{j}^{\left(t\right)}\left(\mathbf{x}_{j},\mathbf{y}_{j}^{\left(t\right)}\right)$
in $\mathcal{P}_{x_{j}}$ with a simple quadratic surrogate objective
function \cite{SSCA} as shown in Fig. \ref{SCA}:
\begin{equation}
\overline{f}_{x_{j}}^{\left(t\right)}\left(\mathbf{x}_{j}\right)\triangleq\left(\mathbf{\boldsymbol{{\rm f}}}_{x_{j}}^{\left(t\right)}\right)^{T}\left(\mathbf{x}_{j}-\mathbf{x}_{j}^{\left(t\right)}\right)+\varGamma_{x_{j}}\left\Vert \mathbf{x}_{j}-\mathbf{x}_{j}^{\left(t\right)}\right\Vert ^{2},\label{eq:surrogate_func_x}
\end{equation}
and obtain an intermediate variable $\overline{\mathbf{x}}_{j}^{\left(t\right)}$
by minimizing the surrogate objective function as
\begin{align}
\overline{\mathbf{x}}_{j}^{\left(t\right)} & =\arg\min\overline{f}_{x_{j}}^{\left(t\right)}\left(\mathbf{x}_{j}\right)\label{eq:px}\\
\text{s.t.} & \quad\hat{\boldsymbol{\Lambda}}_{j}-\Delta\hat{\boldsymbol{\Lambda}}_{j}/2\le x_{j,p}\le\hat{\boldsymbol{\Lambda}}_{j}+\Delta\hat{\boldsymbol{\Lambda}}_{j}/2,\forall p,\nonumber
\end{align}
where $\varGamma_{x_{j}}$ can be any positive number, $\mathbf{\boldsymbol{{\rm f}}}_{x_{j}}^{\left(t\right)}$
is an unbiased estimator of the gradient $\nabla_{\mathbf{x}_{j}}\boldsymbol{L}_{j}^{\left(t\right)}\left(\mathbf{x}_{j},\mathbf{y}_{j}^{\left(t\right)}\right)$,
which is updated recursively as follows:
\begin{equation}
\begin{split}\mathbf{\boldsymbol{{\rm f}}}_{x_{j}}^{\left(t\right)} & =\left(1-\rho^{\left(t\right)}\right)\mathbf{\boldsymbol{{\rm f}}}_{x_{j}}^{\left(t-1\right)}+\frac{\rho^{\left(t\right)}}{B}\sum_{b=1}^{B}\nabla_{\mathbf{x}_{j}}g_{j}^{\left(t\right)}\left(\mathbf{x}_{j}^{\left(t\right)},\mathbf{y}_{j}^{\left(t\right)};\boldsymbol{\Lambda}_{\sim j}^{\left(b\right)}\right),\end{split}
\label{eq:fx}
\end{equation}
where $\left\{ \boldsymbol{\Lambda}_{\sim j}^{\left(b\right)},b=1,...,B\right\} $
is a mini-batch of $B$ samples generated by the distribution $\prod_{j^{'}\neq j}q\left(\boldsymbol{\Lambda}_{j^{'}};\mathbf{x}_{j^{'}},\mathbf{y}_{j^{'}}\right)$
with $\mathbf{x}_{j^{'}}=\mathbf{x}_{j^{'}}^{\left(t+1\right)},\mathbf{y}_{j^{'}}=\mathbf{y}_{j^{'}}^{\left(t+1\right)},\forall j^{'}<j$
and $\mathbf{x}_{j^{'}}=\mathbf{x}_{j^{'}}^{\left(t\right)},\mathbf{y}_{j^{'}}=\mathbf{y}_{j^{'}}^{\left(t\right)},\forall j^{'}>j$,
and $\rho^{\left(t\right)}$ is a decreasing step size that will be
discussed later and we set $\mathbf{\boldsymbol{{\rm f}}}_{x_{j}}^{\left(-1\right)}=\boldsymbol{0}$.
Finally, the updated $\mathbf{x}_{j}$ is given by
\begin{equation}
\mathbf{x}_{j}^{\left(t+1\right)}=\left(1-\gamma^{\left(t\right)}\right)\mathbf{x}_{j}^{\left(t\right)}+\gamma^{\left(t\right)}\overline{\mathbf{x}}_{j}^{\left(t\right)},\label{eq:x_update}
\end{equation}
where $\gamma^{\left(t\right)}$ is another decreasing step size that
will be discussed later, and there is a closed-form solution for $\overline{\mathbf{x}}_{j}^{\left(t\right)}\triangleq\left[\overline{x}_{j,1}^{\left(t\right)},\overline{x}_{j,2}^{\left(t\right)},...,\overline{x}_{j,N_{p}}^{\left(t\right)}\right]^{T}$:
\begin{equation}
\overline{x}_{j,p}^{\left(t\right)}=\left\{ \begin{array}{c}
\hat{\boldsymbol{\Lambda}}_{j}-\Delta\hat{\boldsymbol{\Lambda}}_{j}/2,\quad\qquad\qquad\widetilde{x}_{j,p}^{\left(t\right)}<\hat{\boldsymbol{\Lambda}}_{j}-\Delta\hat{\boldsymbol{\Lambda}}_{j}/2,\\
\widetilde{x}_{j,p}^{\left(t\right)},\quad\quad\hat{\boldsymbol{\Lambda}}_{j}-\Delta\hat{\boldsymbol{\Lambda}}_{j}/2\leq\widetilde{x}_{j,p}^{\left(t\right)}\leq\hat{\boldsymbol{\Lambda}}_{j}+\Delta\hat{\boldsymbol{\Lambda}}_{j}/2,\\
\hat{\boldsymbol{\Lambda}}_{j}+\Delta\hat{\boldsymbol{\Lambda}}_{j}/2,\quad\qquad\qquad\widetilde{x}_{j,p}^{\left(t\right)}>\hat{\boldsymbol{\Lambda}}_{j}+\Delta\hat{\boldsymbol{\Lambda}}_{j}/2,
\end{array}\right.
\end{equation}
where $p=1,...,N_{p}$, $\mathbf{\widetilde{x}}_{j}^{\left(t\right)}\triangleq\left[\widetilde{x}_{j,1}^{\left(t\right)},\widetilde{x}_{j,2}^{\left(t\right)},...,\widetilde{x}_{j,N_{p}}^{\left(t\right)}\right]^{T}=\mathbf{x}_{j}^{\left(t\right)}-\frac{1}{2\Gamma_{x_{j}}}\mathbf{\boldsymbol{{\rm f}}}_{x_{j}}^{\left(t\right)}$.

Similarly, in the $t$-th iteration, the $\left(2j\right)$-th block
$\mathbf{y}_{j}$ is updated by solving the following subproblem:
\[
\begin{split}\mathcal{P}_{y_{j}}:\mathop{\min}\limits _{\mathbf{y}_{j}} & \boldsymbol{L}_{j}^{\left(t\right)}\left(\mathbf{x}_{j}^{\left(t+1\right)},\mathbf{y}_{j}\right)\\
\text{s.t.}\quad & \sum\limits _{p=1}^{N_{p}}y_{j,p}=1,\quad\epsilon\leq y_{j,p}\leq1,\quad\forall p.
\end{split}
\]
In stead of solving $\mathcal{P}_{y_{j}}$ directly, we first construct
a simple quadratic surrogate objective function
\begin{equation}
\overline{f}_{y_{j}}^{\left(t\right)}\left(\mathbf{y}_{j}\right)\triangleq\left(\mathbf{\boldsymbol{{\rm f}}}_{y_{j}}^{\left(t\right)}\right)^{T}\left(\mathbf{y}_{j}-\mathbf{y}_{j}^{\left(t\right)}\right)+\varGamma_{y_{j}}\left\Vert \mathbf{y}_{j}-\mathbf{y}_{j}^{\left(t\right)}\right\Vert ^{2},\label{eq:surrogate_func_y}
\end{equation}
and obtain an intermediate variable $\overline{\mathbf{y}}_{j}^{\left(t\right)}$
by minimizing the surrogate objective function as
\begin{align}
\overline{\mathbf{y}}_{j}^{\left(t\right)} & =\arg\min\overline{f}_{y_{j}}^{\left(t\right)}\left(\mathbf{y}_{j}\right)\label{eq:py}\\
s.t. & \quad\sum\limits _{p=1}^{N_{p}}y_{j,p}=1,\quad\epsilon\leq y_{j,p}\leq1,\quad\forall p,\nonumber
\end{align}
where $\varGamma_{y_{j}}$ can be any positive number, $\mathbf{\boldsymbol{{\rm f}}}_{y_{j}}^{\left(t\right)}$
is an unbiased estimator of the gradient $\nabla_{\mathbf{y}_{j}}\boldsymbol{L}_{j}^{\left(t\right)}\left(\mathbf{x}_{j}^{\left(t+1\right)},\mathbf{y}_{j}^{\left(t\right)}\right)$,
which is updated recursively as follows
\begin{equation}
\begin{split}\mathbf{\boldsymbol{{\rm f}}}_{y_{j}}^{\left(t\right)} & =\left(1-\rho^{\left(t\right)}\right)\mathbf{\boldsymbol{{\rm f}}}_{y_{j}}^{\left(t-1\right)}+\frac{\rho^{\left(t\right)}}{B}\sum_{b=1}^{B}\nabla_{\mathbf{y}_{j}}g_{j}^{\left(t\right)}\left(\mathbf{x}_{j}^{\left(t+1\right)},\mathbf{y}_{j}^{\left(t\right)};\boldsymbol{\Lambda}_{\sim j}^{\left(b\right)}\right)\end{split}
\label{eq:fy}
\end{equation}
and we set $\mathbf{\boldsymbol{{\rm f}}}_{y_{j}}^{\left(-1\right)}=\boldsymbol{0}$.
Finally, the updated $\mathbf{y}_{j}$ is given by
\begin{equation}
\mathbf{y}_{j}^{\left(t+1\right)}=\left(1-\gamma^{\left(t\right)}\right)\mathbf{y}_{j}^{\left(t\right)}+\gamma^{\left(t\right)}\overline{\mathbf{y}}_{j}^{\left(t\right)}.\label{eq:y_update}
\end{equation}

To ensure the convergence of the algorithm, the step sizes $\rho^{\left(t\right)}$
and $\gamma^{\left(t\right)}$ must satisfy the following conditions.

\begin{assumption}[\textit{Assumptions on step sizes}]\label{asm:convstep}$\:$
\begin{enumerate}
\item $\rho^{t}\rightarrow0$, $\sum_{t}\rho^{t}=\infty$, $\sum_{t}\left(\rho^{t}\right)^{2}<\infty$,
\item $\underset{t\rightarrow\infty}{\lim}\gamma^{t}/\rho^{t}=0$.
\end{enumerate}
\end{assumption}

A typical choice of $\rho^{t},\gamma^{t}$ that satisfies Assumption
1 is $\rho^{t}=\mathcal{O}\left(t^{-\kappa_{1}}\right)$, $\gamma^{t}=\mathcal{O}\left(t^{-\kappa_{2}}\right)$,
where $0.5<\kappa_{1}<\kappa_{2}\leq1$. Such form of step sizes have
been widely considered in stochastic optimization \cite{structured_surrogate}.

As can be seen, the surrogate optimization problems in (\ref{eq:px})
and (\ref{eq:py}) are quadratic programming with linear constraints,
which is easy to solve. In addition, explicit expressions of the gradient
$\nabla_{\mathbf{x}_{j}}g_{j}^{\left(t\right)}\left(\mathbf{x}_{j},\mathbf{y}_{j},\boldsymbol{\Lambda}_{\sim j}^{b}\right)$
and $\nabla_{\mathbf{y}_{j}}g_{j}^{\left(t\right)}\left(\mathbf{x}_{j},\mathbf{y}_{j},\boldsymbol{\Lambda}_{\sim j}^{b}\right)$
in (\ref{eq:fx}) and (\ref{eq:fy}) are given in the Appendix \ref{subsec:Grad-Particle-Pos}
and \ref{subsec:Grad-Particle-Wei}, respectively.

The proposed SPVBI is guaranteed to converge to stationary points
of the original Problem $\mathcal{P}_{2}$, as will be proved in Section
\ref{subsec:Convergence-Analysis}. After the convergence, the corresponding
discrete distribution of each variable $q\left(\boldsymbol{\Lambda}_{j}\right)$
composed of particles will approximate the marginal posteriori probability
distribution, as illustrated in Fig. \ref{oscillation_v5}. As a result,
we can take the particle position with the highest probability or
the weighted sum of the particles as the final estimate, which are
the approximate MAP estimate and MMSE estimate, respectively.

\subsection{Convergence Analysis\label{subsec:Convergence-Analysis}}

In this section, we will prove the convergence of the SPVBI algorithm.
First, we present a key lemma which gives some important properties
of the surrogate functions.
\begin{lem}
[Properties of the surrogate functions]\label{lem:Properties-of-suR}For
each iteration $t=1,2,\ldots$ and each block $\mathbf{x}_{j},\mathbf{y}_{j},j=1,2,\ldots J$,
we have
\begin{enumerate}
\item $\overline{f}_{x_{j}}^{\left(t\right)}\left(\mathbf{x}_{j}\right)$
and $\overline{f}_{y_{j}}^{\left(t\right)}\left(\mathbf{y}_{j}\right)$
are uniformly strongly convex in $\mathbf{x}_{j}$ and $\mathbf{y}_{j}$,
respectively.
\item For any $\mathbf{x}_{j}\in\mathcal{X}$\textup{ and} $\mathbf{y}_{j}\in\mathcal{Y}$,
the function $\overline{f}_{x_{j}}^{\left(t\right)}\left(\mathbf{x}_{j}\right)$
and $\overline{f}_{y_{j}}^{\left(t\right)}\left(\mathbf{y}_{j}\right)$,
their derivative, and their second order derivative are uniformly
bounded.
\item $\overline{f}_{x_{j}}^{\left(t\right)}\left(\mathbf{x}_{j}\right)$
and $\overline{f}_{y_{j}}^{\left(t\right)}\left(\mathbf{y}_{j}\right)$
are Lipschitz continuous function w.r.t. $\mathbf{x}_{j}$ and $\mathbf{y}_{j}$,
respectively. Moreover,
\begin{equation}
\limsup_{t_{1},t_{2}\rightarrow\infty}\left|\overline{f}_{x_{j}}^{\left(t_{1}\right)}\left(\mathbf{x}_{j}\right)-\overline{f}_{x_{j}}^{\left(t_{2}\right)}\left(\mathbf{x}_{j}\right)\right|\leq B_{x}\left\Vert \mathbf{x}^{\left(t_{1}\right)}-\mathbf{x}^{\left(t_{2}\right)}\right\Vert ,
\end{equation}
\begin{equation}
\limsup_{t_{1},t_{2}\rightarrow\infty}\left|\overline{f}_{y_{j}}^{\left(t_{1}\right)}\left(\mathbf{y}_{j}\right)-\overline{f}_{y_{j}}^{\left(t_{2}\right)}\left(\mathbf{y}_{j}\right)\right|\leq B_{y}\left\Vert \mathbf{y}^{\left(t_{1}\right)}-\mathbf{y}^{\left(t_{2}\right)}\right\Vert ,
\end{equation}
$\forall\mathbf{x}_{j}\in\mathcal{X}$, $\forall\mathbf{y}_{j}\in\mathcal{Y}$,
for some constants $B_{x}>0$ and $B_{y}>0$.
\item Consider a subsequence $\left\{ \mathbf{x}^{\left(t_{i}\right)},\mathbf{y}^{\left(t_{i}\right)}\right\} _{i=1}^{\infty}$
converging to a limit point $\boldsymbol{x}^{*},\mathbf{y}^{*}$.
There exist uniformly differentiable functions $\hat{f}_{x_{j}}\left(\mathbf{x}_{j}\right)$
and $\hat{f}_{y_{j}}\left(\mathbf{y}_{j}\right)$ such that
\begin{equation}
\mathop{\lim}\limits _{i\to\infty}\overline{f}_{x_{j}}^{\left(t_{i}\right)}\left(\mathbf{x}_{j}\right)=\hat{f}_{x_{j}}\left(\mathbf{x}_{j}\right),\forall\mathbf{x}_{j}\in\mathcal{X},\label{eq:lemma2-4-1-x}
\end{equation}
\begin{equation}
\mathop{\lim}\limits _{i\to\infty}\overline{f}_{y_{j}}^{\left(t_{i}\right)}\left(\mathbf{y}_{j}\right)=\hat{f}_{y_{j}}\left(\mathbf{y}_{j}\right),\forall\mathbf{y}_{j}\in\mathcal{Y}.\label{eq:lemma2-4-1-y}
\end{equation}
Moreover, we have
\begin{equation}
\left\Vert \nabla_{\mathbf{x}_{j}}\hat{f}_{x_{j}}\left(\mathbf{x}_{j}^{*}\right)-\nabla_{\mathbf{x}_{j}}L\left(\mathbf{x}^{*},\mathbf{y}^{*}\right)\right\Vert =0,\label{eq:lemma2-4-2-x}
\end{equation}
\begin{align}
\left\Vert \nabla_{\mathbf{y}_{j}}\hat{f}_{y_{j}}\left(\mathbf{y}_{j}^{*}\right)-\nabla_{\mathbf{y}_{j}}L\left(\mathbf{x}^{*},\mathbf{y}^{*}\right)\right\Vert  & =0.\label{eq:lemma2-4-2-y}
\end{align}
\end{enumerate}
\end{lem}

Please refer to Appendix \ref{subsec:Proof-of-Lemma-2} for the proof.
With the Lemma \ref{lem:Properties-of-suR}, we are ready to prove
the following main convergence result.
\begin{thm}
[Convergence of SPVBI]\label{thm:Convergence-of-SPVBI}Starting
from a feasible initial point $\mathbf{x}^{\left(0\right)},\mathbf{y}^{\left(0\right)}$,
let $\left\{ \mathbf{x}^{\left(t\right)},\mathbf{y}^{\left(t\right)}\right\} _{t=1}^{\infty}$
denote the iterates generated by Algorithm 2. Then every limiting
point $\boldsymbol{x}^{*},\mathbf{y}^{*}$ of $\left\{ \mathbf{x}^{\left(t\right)},\mathbf{y}^{\left(t\right)}\right\} _{t=1}^{\infty}$
is a stationary point of optimization problem $\mathcal{P}$ almost
surely.
\end{thm}

Please refer to Appendix \ref{subsec:Proof-of-Theorem} for the proof.

\subsection{Summary of the SPVBI Algorithm}

The overall SPVBI algorithm is summarized in Algorithm \ref{alg:Stochastic-Particle-based-VBI}.
The mini-batch size $B$ can be chosen to achieve a trade-off between
the per-iteration complexity and the total number of iterations. Thanks
to the idea of averaging over iterations as in \eqref{eq:fx}, \eqref{eq:x_update},
\eqref{eq:fy} and \eqref{eq:y_update}, a constant value of the mini-batch
size $B$ is usually sufficient to achieve a fast convergence, e.g.,
in the simulations, we set $B=10$. Compared to the conventional PVBI
algorithm which needs to calculate a summation of $N_{p}^{J-1}$ terms
when updating one block in each iteration, the proposed SPVBI only
requires to solve a simple quadratic programming problem which only
involves the calculation of $B$ gradients, where $B$ can be much
smaller than $N_{p}^{J-1}$. Moreover, the addition of updating particle
position allows the algorithm to converge more quickly and flexibly
to stable solutions. As such, the proposed SPVBI algorithm has both
lower per-iteration complexity and faster convergence speed than the
conventional PVBI.

The proposed SPVBI can be viewed as an extension of the SSCA framework
\cite{SSCA} in terms of block-wise updating and control-dependent
random state. First, the SSCA in \cite{SSCA} constructs a single
surrogate function to update all the variables simultaneously in each
iteration, while the SPVBI allows block-wise update, which is often
used in VBI-type algorithms due to that a proper block partition facilitates
algorithm derivation and possibly faster convergence \cite{VBI+IS}.
Second, the distribution of the random state in the original SSCA
framework is assumed to be independent of the optimization variable,
i.e., is control independent, while the distribution of the random
state in the SPVBI is control dependent. For example, the random state
$\boldsymbol{\Lambda}_{{\rm \sim}j}$ in the $t$-th iteration follows
the distribution $\prod_{j^{'}\neq j}q\left(\boldsymbol{\Lambda}_{j^{'}};\mathbf{x}_{j^{'}},\mathbf{y}_{j^{'}}\right)$,
which depends on the value of the current optimization variables and
is changing over iterations. Nevertheless, convergence can still be
guaranteed.

\begin{algorithm}[tbh]
\caption{\label{alg:Stochastic-Particle-based-VBI}Stochastic Particle-based
VBI Algorithm}

\textbf{Input}: received signals $\boldsymbol{r}$, multi-band frequency
settings, the estimated number of scattering paths $\hat{K}$, $p\left(\hat{\boldsymbol{\Lambda}}\right)$,
$\left\{ \rho^{\left(t\right)},\gamma^{\left(t\right)}\right\} $.

\textbf{Initialization}: $\left\{ x_{j,p}^{\left(0\right)}\backsim p\left(\hat{\boldsymbol{\Lambda}}_{j}\right),y_{j,p}^{\left(0\right)}=\frac{1}{N_{p}}\right\} _{p=1}^{N_{p}},\forall j$.

\textbf{While not converge do} $(t\rightarrow$$\infty)$

$\quad\quad$\textbf{For} $j=1:J$

$\quad\quad\quad$Generate a mini-batch of realization based on the

$\quad\quad\quad$$\left\{ x_{i,p},y_{i,p}\right\} _{p=1}^{N_{p}},i\ne j$;

$\quad\quad\quad$Update $\mathbf{\boldsymbol{{\rm f}}}_{x_{j}}^{\left(t\right)}$
or $\mathbf{\boldsymbol{{\rm f}}}_{y_{j}}^{\left(t\right)}$;

$\quad\quad\quad$Solving surrogate optimization problem $\mathcal{P}_{x_{j}}$
or $\mathcal{P}_{y_{j}}$;

$\quad\quad\quad$Update the particle position and weight of variable
$\boldsymbol{\Lambda}_{j}$;

$\quad\quad$\textbf{end}

\textbf{end}

\textbf{Output}: Find the position $\tilde{x}_{j},j=1,2,\ldots J$
of the particle with the maximum weight $y_{j}^{\left(max\right)}$.
(i.e. MAP estimation)
\end{algorithm}

\section{Numerical Simulation and Performance Analysis}

In this section, simulations are conducted to demonstrate the performance
of the proposed algorithm. We compare the proposed algorithm with
the following baseline algorithms:
\begin{enumerate}
\item \textbf{Baseline 1} (Root-MUSIC, R-MUSIC) \cite{UWB}: The conventional
root MUSIC algorithm has relatively high accuracy in delay estimation,
so it is adopted here for single-band data, mainly to show the gain
brought by combining the results of multiple bands.
\item \textbf{Baseline 2} (Weighted Root-MUSIC, WR-MUSIC) \cite{WR-MUSIC}:
The WR-MUSIC algorithm is adopted in the coarse estimation stage.
See Section \ref{subsec:Coarse-Estimation} for relevant procedures.
\item \textbf{Baseline 3} (Spectral Estimation, SE) \cite{UWB}: The parameters
of the approximate all-pole model are estimated by spectral search
methods such as root MUSIC algorithm, and the unknown frequency band
data is interpolated according to the model to improve the resolution.
\item \textbf{Baseline 4} (Sparse Bayesian Learning, SBL) \cite{SBL+dictionary}:
The information required for coherent compensation can be extracted
from the SBL solution. Then, by interpolating data between non-contiguous
bands, more accurate estimate can be obtained, where the number of
atoms $M_{ob}$ in the dictionary is set to be $2000$.
\end{enumerate}
In the proposed SPVBI, each variable is equipped with $10$ particles,
and the size of mini-batch $B$ is $10$. The step size sequence is
set as follows: $\rho^{\left(t\right)}=5/\left(5+t\right)^{0.9},\rho^{\left(0\right)}=1$;
$\gamma^{\left(t\right)}=5/\left(15+t\right)^{1},\gamma^{\left(0\right)}=1$.
Unless otherwise specified, the experiment was repeated $400$ times.

In the simulations, we consider both small-bandwidth and large-bandwidth
scenarios, which can cover common WiFi applications such as localization
and imaging. We first describe the common setup for both scenarios.
There are two scattering paths. The amplitude of $\alpha_{k}$ is
$1$ and $0.5$, and $\beta_{k}$ is $-\pi/4$ and $\pi/4$, respectively.
In addition, the phase of $\alpha_{k}$ and initial phase $\phi_{m}$
are uniformly generated within $[0,2\pi]$. The subcarrier spacing
is $78.125$KHz.

In the small-bandwidth scenario, the received signals come from two
non-adjacent frequency bands with a bandwidth of $40$MHz. The initial
frequency is set to $2.4$GHz and $2.52$GHz, respectively. There
are two scattering paths with delays of $25$ns and $500$ns. The
timing synchronization error $\delta_{m}$ is generated following
a Gaussian distribution $\mathcal{N}\left(0,0.01\text{ns}^{2}\right)$.

In the large-bandwidth scenario, the bandwidth of the two non-contiguous
frequency bands is $320$MHz. The initial frequency is set to $5$GHz
and $6$GHz, respectively. The delays of the two scattering paths
are $10$ns and $20$ns, respectively. The timing synchronization
error $\delta_{m}$ is generated following a Gaussian distribution
$\mathcal{N}\left(0,1\text{ns}^{2}\right)$.

\subsection{Performance Comparison with Conventional PVBI Algorithm}

In this section, we will compare the performance of the proposed SPVBI
algorithm with that of the conventional PVBI algorithm.

Firstly, we will focus on the convergence performance. The benefits
of updating particle positions are first shown by comparing the proposed
SPVBI with and without updating particle positions (i.e., PVBI).
\begin{figure}[htbp]
\begin{centering}
\textsf{\includegraphics[scale=0.45]{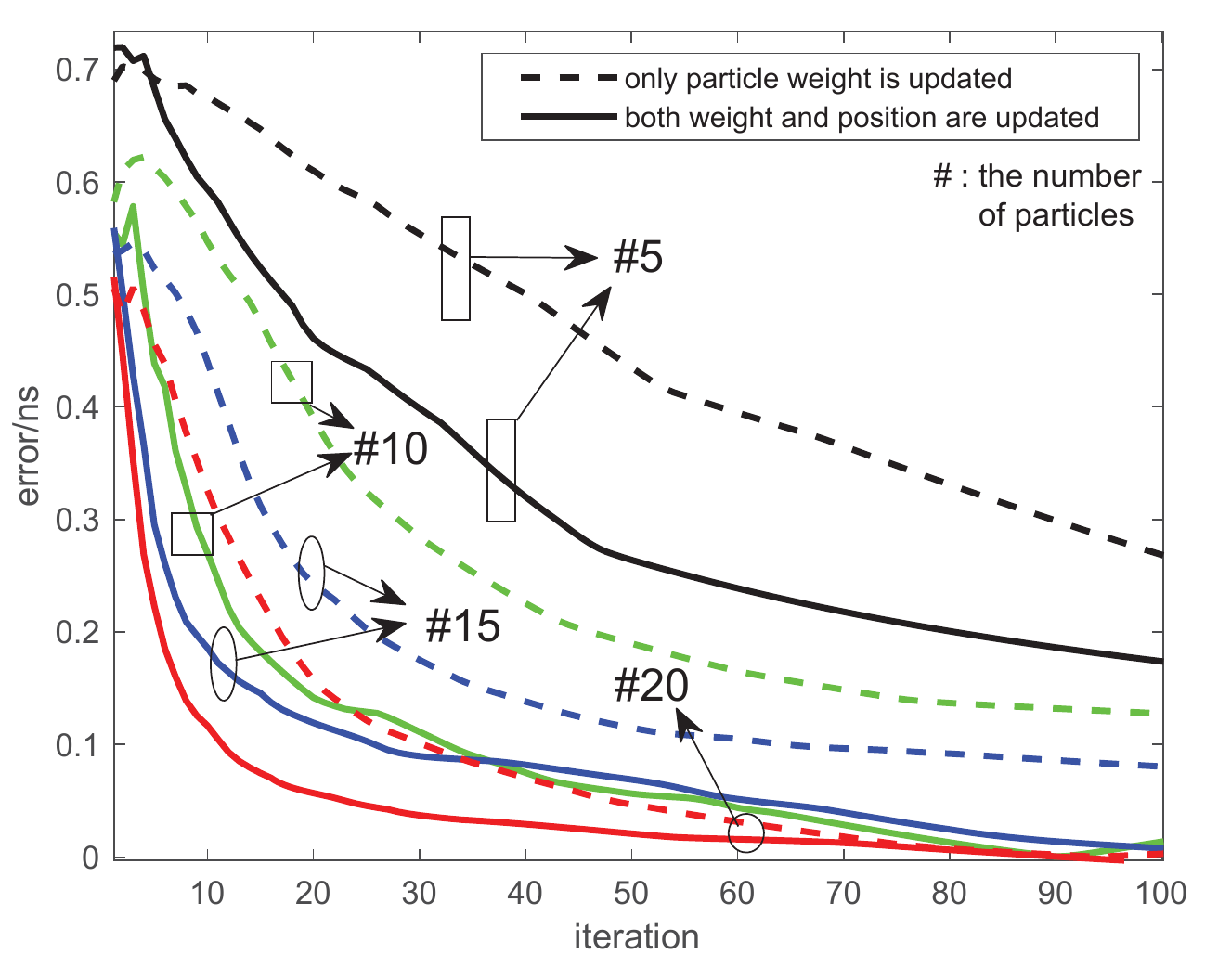}}
\par\end{centering}
\caption{\textsf{\label{iteration_v3}}Iteration error curves with different
particle numbers.}
\end{figure}

In Fig. \ref{iteration_v3}, we plot the convergence curves of different
updating modes with different number of particles under the small-bandwidth
setup. As the number of particles decreases from $20$ to $5$, it
can be seen that the convergence speed will slow down when only the
weight is updated, but the convergence speed and performance will
almost remain the same when the position is also updated. This means
that updating the position of particles can effectively increase the
degree of freedom of optimization and ensure convergence and performance
even with fewer particles.

Secondly, we will focus on the estimation accuracy performance. As
mentioned in the above, due to the large number of variables in multi-band
WiFi sensing scenarios, PVBI algorithms with excessive computational
complexity are not applicable. Therefore, we consider a simplified
case to compare the performance of SPVBI, PVBI, and other algorithms,
where there is only one major scattering center with delay of $50$ns,
and there is no imperfect factors. The SNR is $12$dB, the amplitude
of $\alpha_{1}$ is $1$ and $\beta_{1}$ is $-\pi/4$. The initial
frequencies of the two non-adjacent bands are $2.4$GHz and $2.94$GHz,
respectively, and the bandwidth as well as the frequency step are
$40$MHz and $78.125$KHz, respectively.
\begin{figure}[tbh]
\begin{centering}
\includegraphics[clip,scale=0.45]{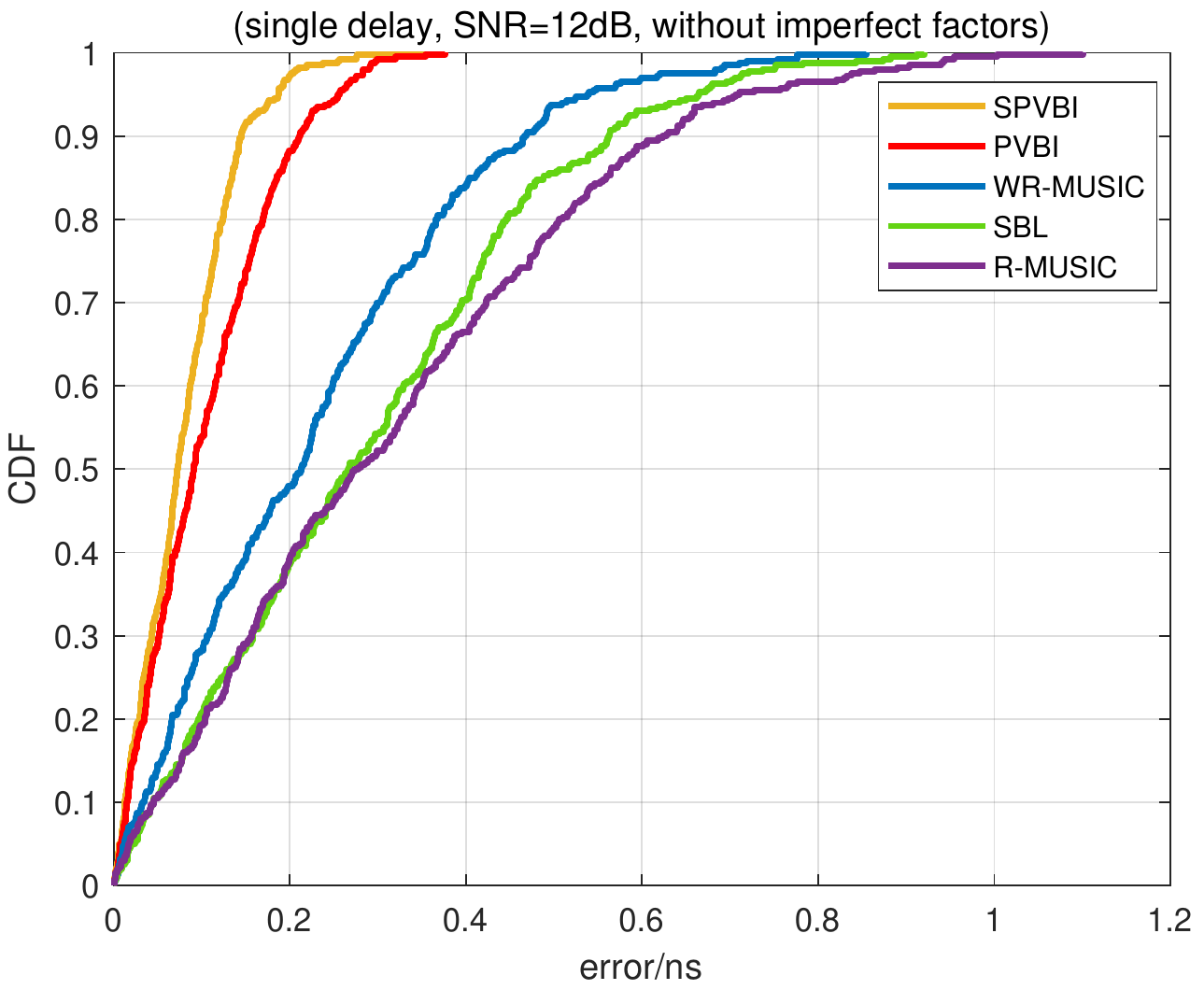}
\par\end{centering}
\caption{\textsf{\label{PVBI}}Performance comparison of the algorithms in
a simplified case.}
\end{figure}

From the error cumulative distribution function (CDF) curve of the
time delay shown in Fig. \ref{PVBI}, it can be seen that SPVBI performs
better than the conventional PVBI, because it additionally update
the position of the particle, so that the particles have a higher
degree of freedom to approximate the posteriori probability. Moreover,
the performance of SPVBI and PVBI algorithms is better than that of
the other algorithms\footnote{SE algorithm is designed to improve the resolution by reconstructing
the full-band data based on the estimated parameters from the R-MUSIC
algorithm, and its delay estimation accuracy is consistent with R-MUSIC.}, because these two VBI-based algorithms can take advantage of the
performance gain brought by multi-band.

In the following, we further consider more general scenarios with
multipath and imperfect factors The detailed parameter settings are
listed at the beginning of Section V.

\subsection{Target Parameter Estimation Error}

In small-bandwidth positioning scenarios, our primary concern is the
distance/range of a certain target. Therefore, we need to investigate
the average delay estimation error and the various factors that can
affect it.

\subsubsection*{Impact of the SNR}

\begin{figure}[htbp]
\begin{centering}
\textsf{\includegraphics[scale=0.45]{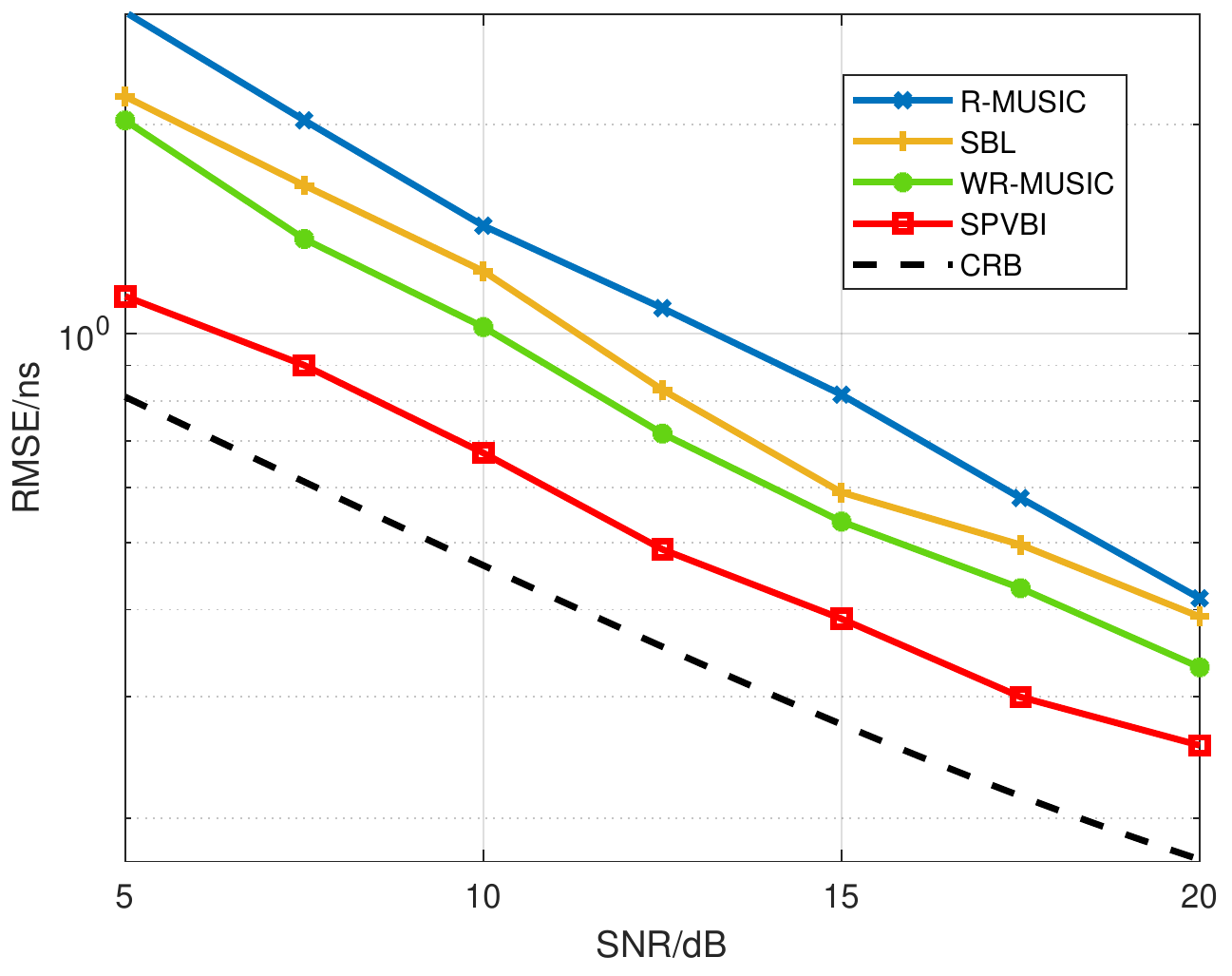}}
\par\end{centering}
\caption{\label{RMSE_SNR}RMSE of delay estimation with respect to the SNR.}
\end{figure}

In Fig. \ref{RMSE_SNR}, it can be seen that the performance of all
algorithms improves as the SNR increases from $5$dB to $20$dB. At
a certain SNR, the root-mean-square error (RMSE) of SPVBI is closer
to the CRB and significantly lower than that of other algorithms,
indicating higher time delay estimation accuracy. The performance
of the SBL algorithm for delay estimation is inferior to WR-MUSIC,
even when the dictionary size is already quite large (i.e. $2000$
atoms).

\subsubsection*{Impact of the band gap}

This subsection presents the impact of the band gap. We changed the
initial frequency of the second band from $2.4$GHz to $2.52$GHz
in steps of $40$MHz to investigate the performance change of the
SPVBI algorithm.
\begin{figure}[htbp]
\begin{centering}
\textsf{\includegraphics[scale=0.45]{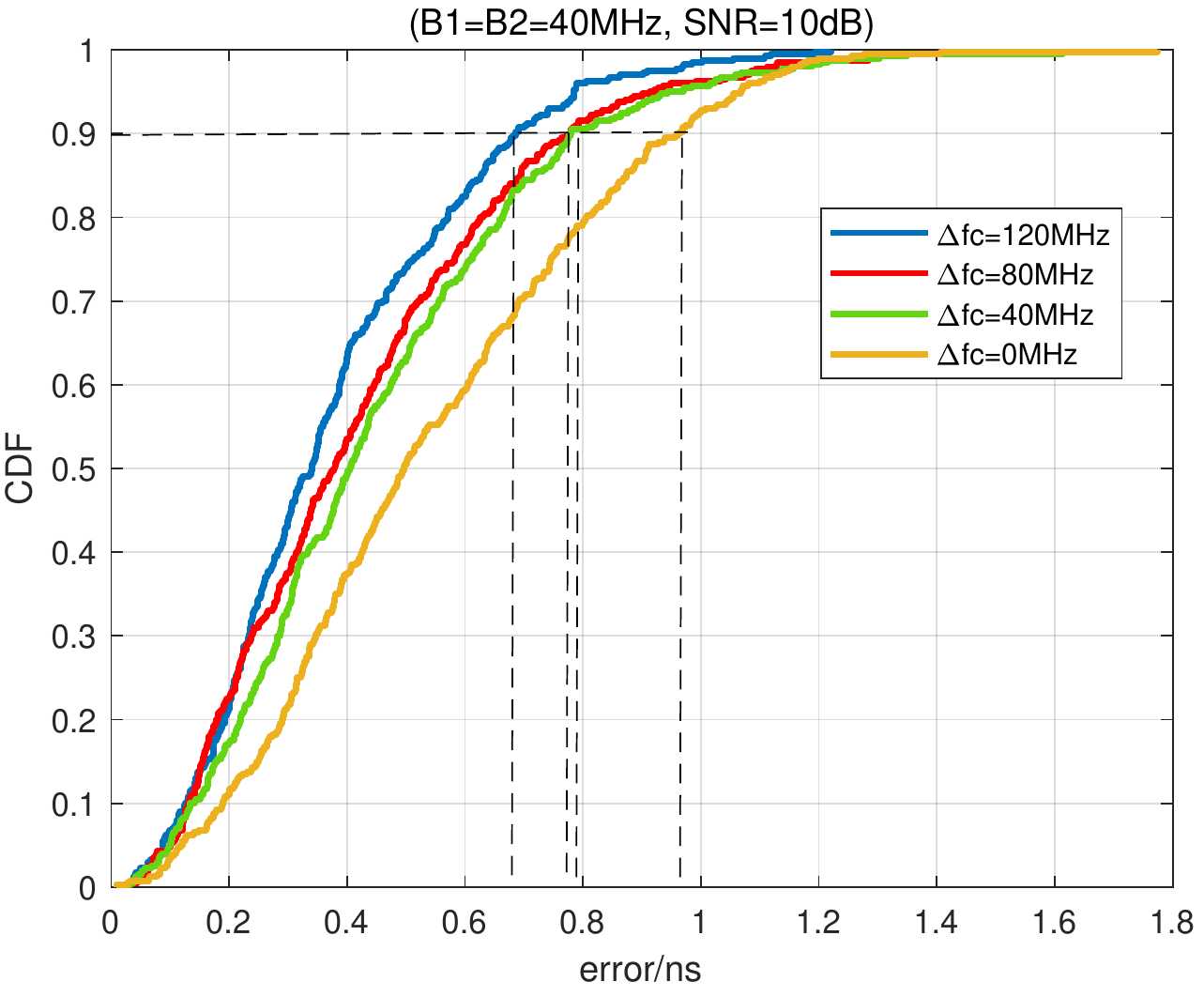}}
\par\end{centering}
\caption{\label{df_freqband}CDF curves of delay estimation errors for different
band gap.}
\end{figure}

In Fig. \ref{df_freqband}, with the increase of frequency band gap,
the performance of SPVBI algorithm keeps improving, which indicates
that the proposed algorithm does utilize the multi-band gap aperture
gain. Intuitively, the larger the frequency band gap, the more obvious
the phase rotation caused by the same delay, which means that the
minor delay variation can also be captured, and hence the better performance.

\subsection{Resolution Performance}

In the application scenarios of large-bandwidth WiFi, such as WiFi
imaging and target feature extraction, ultra-high resolution is required.
In this case, it is often desirable to reconstruct the full-band data
from the available non-adjacent band data to improve the resolution.

However, whether high resolution can be achieved will depend on the
accuracy of the data reconstruction. Therefore, for the SPVBI algorithm
and Baseline $3$ and $4$, we compare the RMSE of data reconstruction
to indirectly show the resolution performance\footnote{The (weighted) root MUSIC algorithm can achieve a good delay estimation
accuracy, but it cannot obtain a good estimation of all parameters
to reconstruct the full-band data. Therefore, we do not compare with
Baseline 1 and 2 in the large-bandwidth scenario.}.

The RMSE between the estimated full-band data and the true full-band
data can be calculated via the following equation: $RMSE_{data}=\sqrt{\frac{1}{N}\sum_{n=1}^{N}\left|r^{\left(n\right)}-\hat{r}^{\left(n\right)}\right|^{2}}$,
where $N$ indicates the number of subcarriers in the full-band.
\begin{figure}[htbp]
\begin{centering}
\textsf{\includegraphics[scale=0.45]{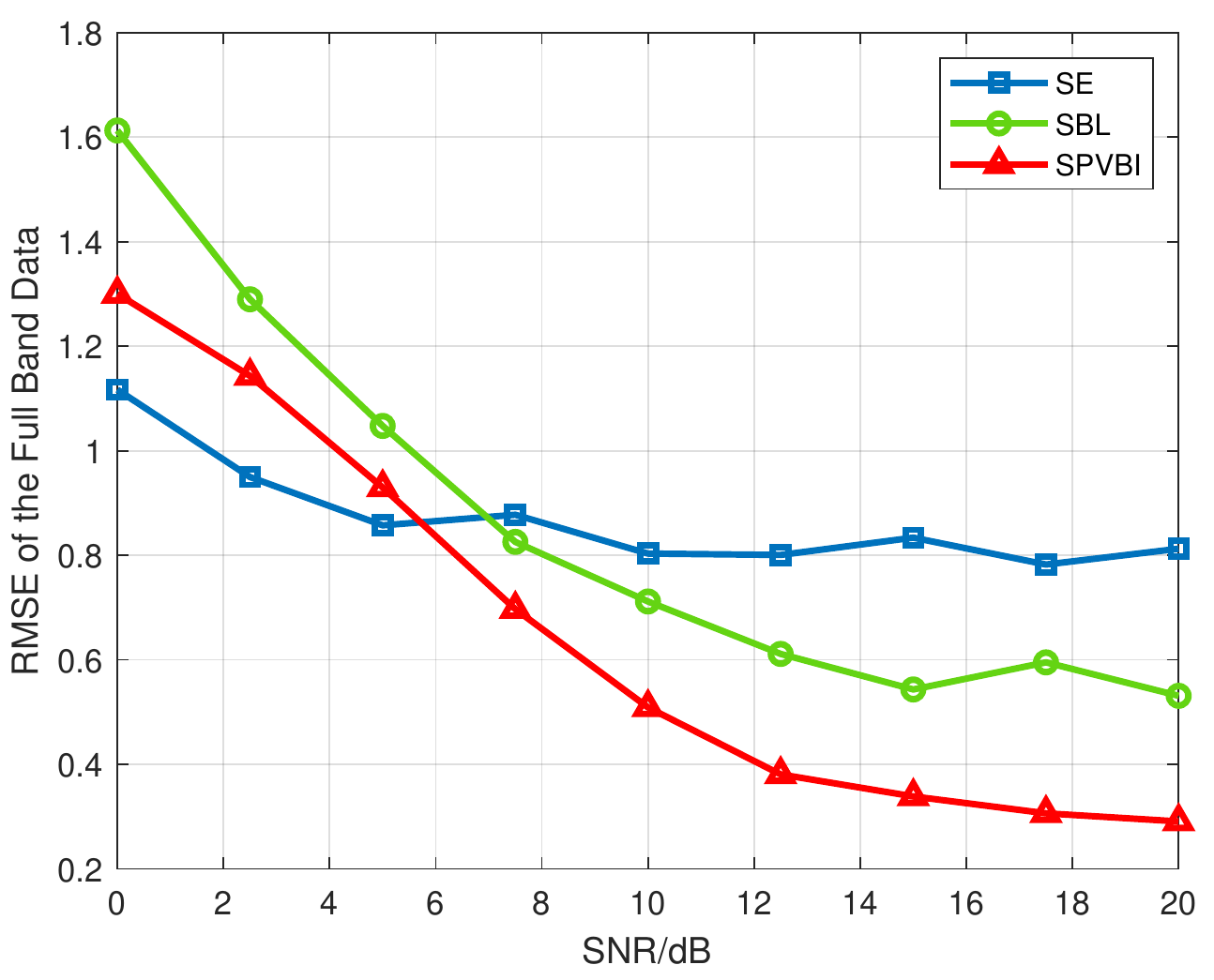}}
\par\end{centering}
\caption{\label{RMSE}RMSE of data reconstruction under different SNR.}
\end{figure}

In Fig. \ref{RMSE}, we show the RMSE of full-band data reconstruction
under different SNR. In the case of low SNR, SE algorithm is less
affected by noise due to its simple model and few parameters to be
estimated. However, with the increase of SNR, compared with other
multi-band fusion algorithms, the full-band data reconstructed by
SPVBI is more accurate, which implies that the proposed algorithm
can obtain more accurate estimation of the signal parameters. The
RMSE performance of SBL algorithm is worse than that of SPVBI and
its complexity is higher.
\begin{figure}[htbp]
\begin{centering}
\textsf{\includegraphics[scale=0.45]{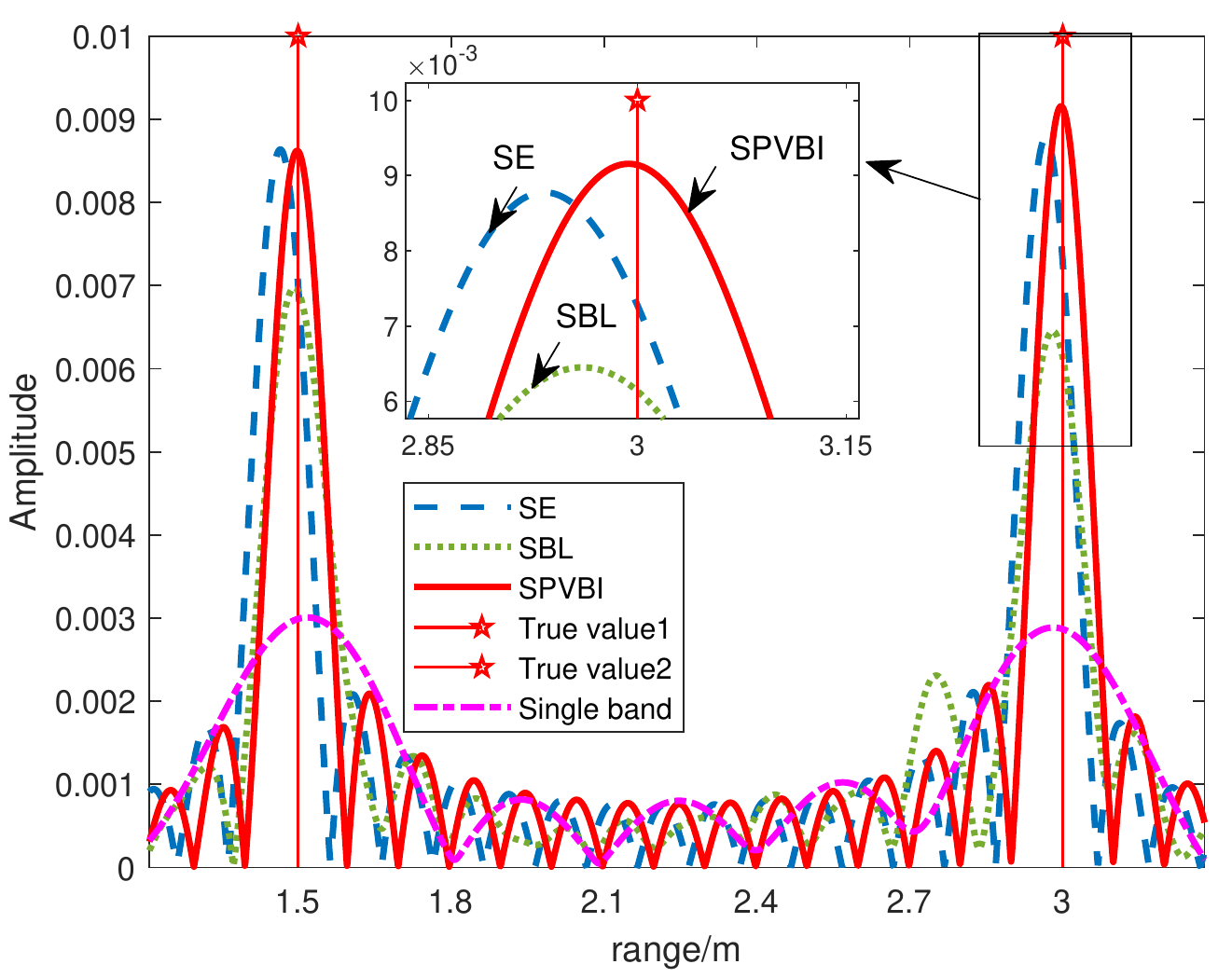}}
\par\end{centering}
\caption{\label{HRRP}High resolution range profile (12dB).}
\end{figure}

In addition, Fig. \ref{HRRP} shows the high resolution range profile
(HRRP) reconstructed by full-band data of different algorithms. It
can be found that the HRRP of the multi-band fusion algorithm (i.e.
SE, SBL, SPVBI) is narrower than that of the single-band reconstruction,
so the resolution is higher. Meanwhile, the RMSE of the full-band
data reconstructed by SPVBI algorithm is smaller, so the peak point
of SPVBI is closer to the true value as shown in the rectangular box.

\subsection{Sensitivity Analysis of the Number of Scattering Paths}

In this subsection, we will focus on the impact of missed detection
and false alarm on the performance for our algorithm, which often
occur in the detection phase and bring the risk of degradation.

\begin{figure}[htbp]
\begin{centering}
\textsf{\includegraphics[clip,scale=0.45]{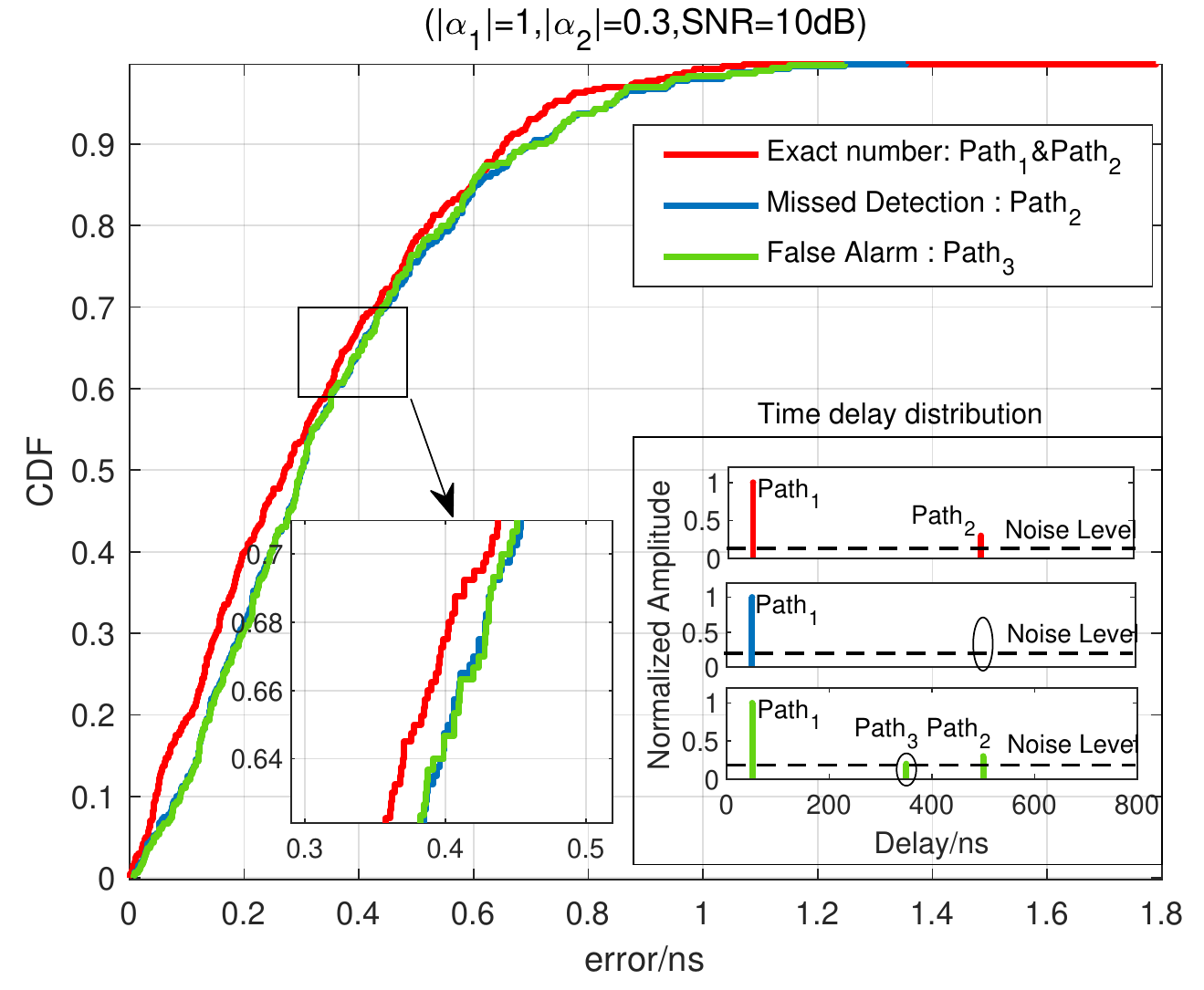}}
\par\end{centering}
\caption{\textsf{\label{mismatch}}Impact of the multipath misdetection on
the performance of SPVBI.}
\end{figure}

Fig. \ref{mismatch} demonstrates the sensitivity of our algorithm
to incorrect paths number at $SNR=10$dB. The red curve represents
the performance of estimating the first path delay when the exact
number is known. The blue and green curve depict cases in which the
path-2 is missed and the false path-3 is detected, respectively. The
amplitudes of $\alpha_{k}$ are $1$ and $0.3$, respectively. The
remaining parameters are the same as in the default simulation setup
in the small-bandwidth scenario.

As shown in Fig. \ref{mismatch}, both missed detection and false
alarm do not significantly degrade the performance of estimating the
first path delay. This indicates that our algorithm is not highly
sensitive to mismatch in the number of scattering paths.

\subsection{Analysis and Comparison of Computational Complexity}

In this subsection, our focus will be on analyzing the complexity
of the different algorithms and presenting the complexity results
of them.

For Root-MUSIC (R-MUSIC) \cite{rao1989root-MUSIC,UWB} algorithm,
it mainly includes subspace decomposition and polynomial rooting,
with the complexity of $\mathcal{O}\left(MN_{m}^{3}\right)$ and $\mathcal{O}\left(M\left(2N_{m}-1\right)^{3}\right)$,
respectively. $M$ is the number of frequency bands and $N_{m}$ denotes
the number of subcarriers in each band. For multi-band weighted root-MUSIC
(WR-MUSIC) algorithm \cite{WR-MUSIC}, the complexity is about $M$
times that of R-MUSIC algorithm. In addition, the computational complexity
of least square (LS) method in the coarse estimation stage is approximately
$\mathcal{O}\left(N_{m}K^{2}+K^{3}\right)$, which is negligible compared
with WR-MUSIC. $K$ is the number of scattering paths. For SE algorithm
\cite{UWB}, in addition to the R-MUSIC algorithm used separately
for each band, there is a nonlinear least square fitting step, which
is solved by Levenberg-Marquarelt (LM) algorithm with a complexity
order of $\mathcal{O}\left(N_{m}^{3}\right)$. Therefore, the total
computational complexity of SE algorithm is $M\cdot\mathcal{O}_{R-MUSIC}+\mathcal{O}\left(N_{m}^{3}\right)$.
For SBL algorithm \cite{SBL+dictionary}, the main cost depends on
the size of the observation matrix used, with a complexity order of
$\mathcal{O}\left(M\cdot M_{ob}^{3}\right)$, $M_{ob}$ is the number
of atoms.

For PVBI algorithm, its per-iteration complexity order is $\mathcal{O}\left(J\cdot\left(N_{p}\right)^{J-1}N_{LH}\right)$,
where $N_{LH}=\mathcal{O}\left(KMN_{m}\right)$ represents the average
number of floating point operations (FLOPs) required to compute the
dominant logarithmic likelihood value. For SPVBI algorithm, through
mini-batch sampling and minimization of quadratic surrogate objective
functions, the per-iteration complexity can be reduced to $\mathcal{O}\left(2J\cdot\left(N_{p}BN_{grad}+N_{p}^{3}\right)\right)$,
where $BN_{grad}$ and $N_{p}^{3}$ represent the complexity of computing
a mini-batch of gradients and quadratic programming search, respectively,
where $N_{grad}=\mathcal{O}\left(KMN_{m}\right)$ represents the average
number of FLOPs required to compute a gradient. It can be seen that
SPVBI greatly reduces the computational complexity compared with conventional
PVBI.

In Table \ref{Order_PerIteration}, we summarize the per-iteration
complexity order of different algorithms, and numerically compare
them under a typical setting as follows: $J=8$, $N_{p}=10$, $M=2$,
$N_{m}=256$, $B=10$, $N_{LH}=20000$, $N_{grad}=18000$, $M_{ob}=2000$.
\begin{table}[tbh]
\caption{\label{Order_PerIteration}COMPARISON OF THE COMPLEXITY ORDER FOR
DIFFERENT ALGORITHMS.}

\centering{}%
\begin{tabular}{|c|c|c|}
\hline
Algorithms & Complexity order per iteration & Typical values\tabularnewline
\hline
PVBI & $\mathcal{O}\left(J\cdot\left(N_{p}\right)^{J-1}\cdot N_{LH}\right)$ & $1.56\times10^{12}$\tabularnewline
\hline
SPVBI & $\mathcal{O}\left(2J\cdot\left(N_{p}BN_{grad}+N_{p}^{3}\right)\right)$ & $2.88\times10^{7}$\tabularnewline
\hline
R-MUSIC & $\mathcal{O}\left(N_{m}^{3}+\left(2N_{m}-1\right)^{3}\right)$ & $1.50\times10^{8}$\tabularnewline
\hline
WR-MUSIC & $\mathcal{O}\left(MN_{m}^{3}+M\left(2N_{m}-1\right)^{3}\right)$ & $3.00\times10^{8}$\tabularnewline
\hline
SE & $M\cdot\mathcal{O}_{R-MUSIC}+\mathcal{O}\left(N_{m}^{3}\right)$ & $3.17\times10^{8}$\tabularnewline
\hline
SBL & $\mathcal{O}\left(M\cdot M_{ob}^{3}\right)$ & $1.60\times10^{10}$\tabularnewline
\hline
\end{tabular}
\end{table}

It can be found that SPVBI algorithm can achieve good trade-off between
performance and complexity. PVBI algorithm can only be applied in
scenarios with few variables, and once the number of variables is
large, the complexity will be unacceptable. Although the computational
complexity of two-stage estimation (i.e. $\mathcal{O}_{WR-MUSIC}+\mathcal{O}_{SPVBI}$)
is higher than that of the WR-MUSIC and SE algorithm, the estimation
accuracy and resolution are greatly improved. Compared with the SBL
algorithm, the complexity of SPVBI is greatly reduced and the performance
is also improved.

\section{Conclusions}

In this paper, we proposed a novel two-stage design for multi-band
WiFi sensing. To overcome the difficulty caused by the oscillation
of likelihood function, we adopt two distinct signal models, which
are transformed from the original multi-band signal model and used
within a two-stage estimation framework. The coarse estimation stage
helps to reduce the computational complexity by narrowing down the
estimation range, and the refined estimation stage leverages the carrier
phase information between different frequency bands (i.e., multi-band
gap gain) to improve estimation performance further. Moreover, the
SPVBI algorithm based on block SSCA transforms the computation of
expectation with exponential complexity in the conventional PVBI into
solving stochastic optimization problems, guaranteeing convergence
theoretically and reducing computational complexity through mini-batch
random sampling and averaging over iterations. Simulation results
indicate that the proposed algorithm achieves good performance with
acceptable complexity across different scenarios. Moreover, adding
the particle position update can speed up convergence and reduce the
required number of particles. It is worth mentioning that the proposed
framework has good generalization ability, and is expected to be applied
to more high-dimensional non-convex parameter estimation scenarios
with numerous local optimums and non-conjugated prior. In the future,
other more expressive methods to characterize the posterior distribution
and the combination with artificial intelligence can also be further
explored.

\appendix

\subsection{Proof of Lemma \ref{lem:Properties-of-suR} \label{subsec:Proof-of-Lemma-2}}

We first introduce the following preliminary result.
\begin{lem}
\label{lem:Properties-of-suR-1}Given subproblem $\mathcal{P}_{x_{j}}$
and $\mathcal{P}_{y_{j}}$ under Lemma 2, suppose that the step sizes
$\rho^{t}$and $\gamma^{t}$ are chosen according to Assumption 1.
Let $\left\{ \mathbf{x}^{\left(t\right)},\mathbf{y}^{\left(t\right)}\right\} $
be the sequence generated by Algorithm 2. Then, the following holds
\begin{equation}
\mathop{\lim}\limits _{t\to\infty}\left|\mathbf{\boldsymbol{{\rm f}}}_{x_{j}}^{\left(t\right)}-\nabla_{\mathbf{x}_{j}}\boldsymbol{L}_{j}^{\left(t\right)}\left(\mathbf{x}_{j}^{\left(t\right)},\mathbf{y}_{j}^{\left(t\right)}\right)\right|=0,w.p.1.
\end{equation}
\begin{equation}
\mathop{\lim}\limits _{t\to\infty}\left|\mathbf{\boldsymbol{{\rm f}}}_{y_{j}}^{\left(t\right)}-\nabla_{\mathbf{y}_{j}}\boldsymbol{L}_{j}^{\left(t\right)}\left(\mathbf{x}_{j}^{\left(t+1\right)},\mathbf{y}_{j}^{\left(t\right)}\right)\right|=0.w.p.1.
\end{equation}
\end{lem}
\begin{IEEEproof}
Lemma \ref{lem:Properties-of-suR-1} is a consequence of (\cite{lemma},
Lemma 1). We only need to verify that all the technical conditions
therein are satisfied. Specifically, Condition (a) of (\cite{lemma},
Lemma 1) is satisfied because $\mathcal{X}$ and $\mathcal{Y}$ are
compact and bounded. Condition (b) of (\cite{lemma}, Lemma 1) follows
from the boundedness of the instantaneous gradient $\nabla g_{j}^{\left(t\right)}$.
Conditions (c)–(d) immediately come from the step-size rule (1) in
Assumption 1. Although the control-dependent random states are not
identically distributed over iterations, the distributions (i.e.,
positions and weights of particles) change slowly at the rate of order
$\mathcal{O}\left(\gamma^{t}\right)$, so we have $\left\Vert \nabla\boldsymbol{L}_{j}^{\left(t+1\right)}-\nabla\boldsymbol{L}_{j}^{\left(t\right)}\right\Vert =\mathcal{O}\left(\gamma^{t}\right)$.
Plusing the step-size rule 2) in Assumption 1, Condition (e) of (\cite{lemma},
Lemma 1) is also satisfied.
\end{IEEEproof}
Using this result, since $\nabla\boldsymbol{L}_{j}^{\left(t\right)}\left(\mathbf{x}_{j}^{\left(t\right)},\mathbf{y}_{j}^{\left(t\right)}\right)$
is obviously bounded, then $\mathbf{\boldsymbol{{\rm f}}}_{x_{j}}^{\left(t\right)}$
and $\mathbf{\boldsymbol{{\rm f}}}_{y_{j}}^{\left(t\right)}$ are
bounded. As can be seen, the surrogate function adopted is a convex
quadratic function with box constraints. Therefore, 1)-3) in Lemma
\ref{lem:Properties-of-suR} follow directly from the expression of
the surrogate function in \eqref{eq:surrogate_func_x} and \eqref{eq:surrogate_func_y}.

For 4) in Lemma \ref{lem:Properties-of-suR}, the proof is similar
to that of (\cite{SSCA}, Lemma 1). Due to 1)-3) in Lemma \ref{lem:Properties-of-suR},
the families of functions $\left\{ \overline{f}_{x_{j}}^{\left(t_{i}\right)}\left(\mathbf{x}_{j}\right)\right\} $
and $\left\{ \overline{f}_{y_{j}}^{\left(t_{i}\right)}\left(\mathbf{y}_{j}\right)\right\} $
are equicontinuous. Moreover, they are bounded and defined over a
compact set $\mathcal{X}$ and $\mathcal{Y}$. Hence the Arzela–Ascoli
theorem \cite{1988Linear} implies that, by restricting to a subsequence,
there exists uniformly continuous functions $\hat{f}_{x_{j}}\left(\mathbf{x}_{j}\right)$
and $\hat{f}_{y_{j}}\left(\mathbf{y}_{j}\right)$ such that \eqref{eq:lemma2-4-1-x}
and \eqref{eq:lemma2-4-1-y} in Lemma \ref{lem:Properties-of-suR}-4)
are satisfied. Clearly, we have
\begin{equation}
\nabla_{\mathbf{x}_{j}/\mathbf{y}_{j}}\boldsymbol{L}\left(\mathbf{x}^{*},\mathbf{y}^{*}\right)=\nabla_{\mathbf{x}_{j}/\mathbf{y}_{j}}\boldsymbol{L}_{j}\left(\mathbf{x}^{*},\mathbf{y}^{*}\right)
\end{equation}
almost surely. And because of $\mathop{\lim}\limits _{i\to\infty}\mathbf{\boldsymbol{{\rm f}}}_{x_{j}}^{\left(t_{i}\right)}=\nabla_{\mathbf{x}_{j}}\hat{f}_{x_{j}}\left(\mathbf{x}_{j}^{*}\right)$,
$\mathop{\lim}\limits _{i\to\infty}\mathbf{\boldsymbol{{\rm f}}}_{y_{j}}^{\left(t_{i}\right)}=\nabla_{\mathbf{y}_{j}}\hat{f}_{y_{j}}\left(\mathbf{y}_{j}^{*}\right)$
and Lemma \ref{lem:Properties-of-suR-1}, we further have
\begin{align}
\left\Vert \nabla_{\mathbf{x}_{j}}\hat{f}_{x_{j}}\left(\mathbf{x}_{j}^{*}\right)-\nabla_{\mathbf{x}_{j}}\boldsymbol{L}\left(\mathbf{x}^{*},\mathbf{y}^{*}\right)\right\Vert  & =0,\\
\left\Vert \nabla_{\mathbf{y}_{j}}\hat{f}_{y_{j}}\left(\mathbf{y}_{j}^{*}\right)-\nabla_{\mathbf{y}_{j}}\boldsymbol{L}\left(\mathbf{x}^{*},\mathbf{y}^{*}\right)\right\Vert  & =0.
\end{align}

\subsection{Proof of Theorem \ref{thm:Convergence-of-SPVBI} \label{subsec:Proof-of-Theorem}}

It is easy to see that each iteration of Algorithm 2 is equivalent
to optimizing the following surrogate function
\begin{equation}
\overline{f}^{\left(t\right)}\left(\mathbf{x},\mathbf{y}\right)=\sum\limits _{j=1}^{J}\left[\overline{f}_{x_{j}}^{\left(t\right)}\left(\mathbf{x}_{j}\right)+\overline{f}_{y_{j}}^{\left(t\right)}\left(\mathbf{y}_{j}\right)\right].\label{eq:whole_surrogate_func}
\end{equation}
Moreover, from Lemma \ref{lem:Properties-of-suR}, we have
\begin{equation}
\mathop{\lim}\limits _{t\to\infty}\overline{f}^{\left(t\right)}\left(\mathbf{x},\mathbf{y}\right)=\hat{f}\left(\mathbf{x},\mathbf{y}\right)\triangleq\sum\limits _{j=1}^{J}\left[\hat{f}_{x_{j}}\left(\mathbf{x}_{j}\right)+\hat{f}_{y_{j}}\left(\mathbf{y}_{j}\right)\right].\label{eq:Grad_AsyConsi_x}
\end{equation}
Using Lemma \ref{lem:Properties-of-suR} and the similar analysis
as in the proof of (\cite{SSCA}, Theorem 1), we have that
\begin{align}
\left\{ \mathbf{x}^{*},\mathbf{y}^{*}\right\}  & =\arg\min\hat{f}\left(\mathbf{x},\mathbf{y}\right)\nonumber \\
\text{s.t. \quad\ensuremath{h_{j}\left(\mathbf{x},\mathbf{y}\right)\leq}0,} & \quad H_{j}\left(\mathbf{x},\mathbf{y}\right)=0,\quad j=1,\ldots,J\label{eq:Prob_ConvSurroFunc}
\end{align}
where $h_{j}\left(\mathbf{x},\mathbf{y}\right)$ and $H_{j}\left(\mathbf{x},\mathbf{y}\right)$
represent inequality constraints and equality constraints in the original
problem, respectively. The Karush-Kuhn-Tucker (KKT) condition of problem
\eqref{eq:Prob_ConvSurroFunc} implies that there exist $\lambda_{1},\ldots,\lambda_{J}$
and $\mu_{1},\ldots,\mu_{J}$ that
\begin{align}
\nabla\hat{f}\left(\mathbf{x}^{*},\mathbf{y}^{*}\right)+\sum\limits _{j=1}^{J}\mu_{j}\nabla h_{j}\left(\mathbf{x}^{*},\mathbf{y}^{*}\right)+\sum\limits _{j=1}^{J}\lambda_{j}\nabla H_{j}\left(\mathbf{x}^{*},\mathbf{y}^{*}\right) & =0\nonumber \\
h_{j}\left(\mathbf{x}^{*},\mathbf{y}^{*}\right)\leq0,\quad\quad H_{j}\left(\mathbf{x}^{*},\mathbf{y}^{*}\right)=0,\quad j=1,\ldots,J\nonumber \\
\mu_{j}\geq0,\quad\quad\mu_{j}h_{j}\left(\mathbf{x}^{*},\mathbf{y}^{*}\right)=0,\quad j=1,\ldots,J.\label{eq:KKT}
\end{align}
Finally, it follows from Lemma \ref{lem:Properties-of-suR} and \eqref{eq:KKT}
that $\left\{ \mathbf{x}^{*},\mathbf{y}^{*}\right\} $ also satisfies
the KKT condition of Problem $\mathcal{P}$. This completes the proof.

\subsection{Derivation of Particle Position Gradient\label{subsec:Grad-Particle-Pos}}

The general formula for the gradient of position is
\begin{align}
 & \frac{\partial g_{j}^{\left(t\right)}\left(\mathbf{x}_{j}^{\left(t\right)},\mathbf{y}_{j}^{\left(t\right)},\boldsymbol{\Lambda}_{\sim j}^{\left(b\right)}\right)}{\partial x_{j,p}^{\left(t\right)}}\nonumber \\
 & =-y_{j,p}^{\left(t\right)}\left[\frac{\partial\ln p\left(x_{j,p}^{\left(t\right)}\right)}{\partial x_{j,p}^{\left(t\right)}}+\frac{\partial\ln p\left(\boldsymbol{r}\left|\boldsymbol{\Lambda}_{\sim j}^{\left(b\right)},x_{j,p}^{\left(t\right)}\right.\right)}{\partial x_{j,p}^{\left(t\right)}}\right],
\end{align}
where the superscript $\left(b\right)$ indicates that its value is
generated by the $b$-th sample of a mini-batch.

Without any additional priors, we assume that the variables are uniformly
distributed in the interval, so the derivatives $\frac{\partial\ln p\left(x\right)}{\partial x}$
defined in the interval are all $0$. For the Gaussian prior $\mathcal{N}\left(0,var\right)$,
the derivative is

\begin{equation}
\frac{\partial\ln p\left(x\right)}{\partial x}=\frac{p\left(x\right)}{p\left(x\right)}\left[\frac{-\left(x-0\right)}{var}\right]=-x/var.
\end{equation}

Next, we will focus on the gradient of the likelihood function with
respect to position:
\begin{align}
\frac{\partial\ln p\left(\boldsymbol{r}\left|\boldsymbol{\Lambda}_{\sim j}^{\left(b\right)},x_{j,p}^{\left(t\right)}\right.\right)}{\partial x_{j,p}^{\left(t\right)}} & =\frac{-1}{{\rm 2}\eta_{w}^{2}}\sum\limits _{m=1}^{M}\sum\limits _{n=0}^{N_{m}-1}\frac{\partial\left|r_{m}^{\left(n\right)}-s_{m}^{\left(n\right)}\right|^{2}}{\partial x_{j,p}^{\left(t\right)}}\nonumber \\
=\frac{-1}{\eta_{w}^{2}}\sum\limits _{m=1}^{M}\sum\limits _{n=0}^{N_{m}-1} & \left[Re\left(s_{m}^{\left(n\right)}-r_{m}^{\left(n\right)}\right)\frac{\partial Re\left(s_{m}^{\left(n\right)}\right)}{\partial x_{j,p}^{\left(t\right)}}\right.\nonumber \\
+Im\left(s_{m}^{\left(n\right)}-r_{m}^{\left(n\right)}\right) & \left.\frac{\partial Im\left(s_{m}^{\left(n\right)}\right)}{\partial x_{j,p}^{\left(t\right)}}\right],
\end{align}
where the terms $\frac{\partial Re\left(s_{m}^{\left(n\right)}\right)}{\partial x_{j,p}^{\left(t\right)}}$
and $\frac{\partial Im\left(s_{m}^{\left(n\right)}\right)}{\partial x_{j,p}^{\left(t\right)}}$
for different variables can be easily derived according to \eqref{eq:reconstruc_sig}.

\subsection{Derivation of Particle Weight Gradient\label{subsec:Grad-Particle-Wei}}

The general formula for the gradient of weight is
\begin{align}
 & \frac{\partial g_{j}^{\left(t\right)}\left(\mathbf{x}_{j}^{\left(t+1\right)},\mathbf{y}_{j}^{\left(t\right)},\boldsymbol{\Lambda}_{\sim j}^{\left(b\right)}\right)}{\partial y_{j,p}^{\left(t\right)}}\nonumber \\
 & =\ln y_{j,p}^{\left(t\right)}-\ln p\left(x_{j,p}^{\left(t+1\right)}\right)-\ln p\left(r\left|\boldsymbol{\Lambda}_{{\rm \sim}j}^{\left(b\right)},x_{j,p}^{\left(t+1\right)}\right.\right)+1.
\end{align}

\bibliographystyle{IEEEtran}
\bibliography{SP-VBI}

\end{document}